\documentclass[11pt,aps,pra,eqsecnum,subeqn,graphicx]{article}
\usepackage{amssymb,amsmath,color,times,epsfig}
\newcommand{\SU}{\rm{SU}}

\newcommand{\ds}{\displaystyle}
\newcommand{\be}{\begin{equation}}
\newcommand{\ee}{\end{equation}}
\newcommand{\ba}{\begin{array}}
\newcommand{\ea}{\end{array}}
\newcommand{\beqa}{\begin{eqnarray}}
\newcommand{\eeqa}{\end{eqnarray}}
\newcommand{\beqas}{\begin{eqnarray*}}
\newcommand{\eeqas}{\end{eqnarray*}}
\newcommand{\beqal}{\begin{lefteqnarray}}
\newcommand{\eeqal}{\end{lefteqnarray}}

\begin{document}

 \thispagestyle{empty}

\title{Charged right-handed Higgsino field contribution to the chargino mass
spectrum and inverse parameter problem  in Left-Right Supersymmetric
Models}

\author{Nibaldo Alvarez-Moraga${}^{1,}$\thanks{{\it Email address:}
nibaldo.alvarez.m@exa.pucv.cl} \; and Artorix de la Cruz de
O\~na${}^{1,2,}$\thanks{{\it Email address:}
artorde@vax2.concordia.ca}
\\ {\small \it  ${}^1$ Autonomous Center of theoretical Physics and Applied Mathematics,}
\\
{\small \it 4905 Des \'erables $\#$ 207, Pierrefonds (Qu\'ebec)
H9J 2W3, Canada.}\\
{\it ${}^2$ \small \it \small D\'epartement de g\'enie physique
Ecole Polytechnique de Montr\'eal,} \\ { \small \it  C.P. 609, succ.
Centre-Ville, (Qu\'ebec) H3C 3A7, Canada.} }

\maketitle

\begin{abstract}
The contribution of the charged right-handed higgsino fields to the
chargino mass spectrum in the context of the Left-Right
Supersymmetric Model is studied. Analytical expressions for the
chargino masses assuming arbitrary CP-violating phases are given.
Also, the corresponding inverse parameter problem is studied.
Analytical disentangling of some relevant parameters is presented. A
general inversion algorithm,  based on suitable measurement of
cross-section type observables associated to chargino pair
production in electron positron annihilation,  is proposed.

\end{abstract}


 \vspace{5mm}

\setcounter{equation}{0}
\newpage
\section{INTRODUCTION}
In most of the Left-Right Supersymmetric (LRSUSY) model, the
right-handed symmetry-breaking energy scale is taken so many order
greater than that of its left-handed counterpart due to the great
right-handed gauge boson masse requirements and the implementation
of the so-called seesaw mechanism \cite{sesaw1,AMRS}. Thus at
electro-weak scales, the charged right-handed fields commonly appear
decoupled from its left-handed counterpart. However, authors
studying left-right symmetric and supersymmetric models, have
demonstrated that a moderate decoupling limit is also possible, by
introducing an intermediate scale or an extra symmetry, and that it
could provide testable effects of the remnants of right-handed
symmetries in upcoming collider experiments
\cite{AMRS,AMS,AMRS2,ABMRS,keyvLvR1}.

In this article, in the context of  on the LRSUSY  model, we compute
the chargino mass spectrum analytically considering the
contributions of the charged right handed Higssino fields. To
determine the chargino masses we must solve a quintic algebraic
equation. However, we will see that the analytical resolution of
this quintic equation is always possible due to the simple
dependence of one of the physical chargino masses with one of the
fundamental Higgsino mass parameters. Next, we study the
corresponding inverse problem, i.e., how to determinate the
fundamental LRSUSY parameters when the physical chargino masses and
some measurable physical obervables are known. At this stage, the
problem is more complex. Nevertheless, as we will see in the last
section of this paper, a complete analytical inversion is in
principle possible using the projector formalism \cite{Jarlskog}.
The projector method used in this paper constitutes an extension of
the formalism used by some authors to compute the neutralino and
chargino mass spectrum, to study the corresponding direct and
inverse parameter problems, in the context of the Minimal
Supersymmetric Standard Model (MSSM) as well as in the context to
the LRSUSY model \cite{key12,key13,key11,nalvar8}.

This paper is organized as follows. In Section \ref{sec-LRSUSY}, we
give a brief description of the LRSUSY model focusing  our attention
on the chargino sector of it. We write the chargino sector
Lagrangian density in terms of the charged Higgsino fields,
including the right-handed ones, and of the  chargino mass matrix.
The chargino mass matrix is expressed in terms of  the fundamental
chargino parameters and an arbitrary set of CP-violating phases. In
Section \ref{sec-chargino-MASS}, we compute the chargino mass
spectrum analytically.  We describe the explicit dependence of the
chargino masses  on the right-handed Higgsino parameters and on the
CP-violating phases.  In Section \ref{sec-diagonalizing}, we compute
the two matrices we need to diagonalize the chargino mass matrix.
 In Section
\ref{sec-projector-formalism} we generate a novel approach of the
projector theory and we  connect this theory with the Jarlskog's
projector formalism.  In Section \ref{sec-det-parameters} we use a
system of basic equations derived from the projector theory to
disentangle some relevant chargino fundamental parameters from the
rest of them. In Section \ref{sec-par-inversion} we analyze the
inverse parameter problem. A conclusion and some comments are
provided in Section \ref{sec-conclusion}. Finally,  explicit
expressions of the entries of the matrices formed from the product
between the chargino mass matrix and its corresponding adjoint
matrix, and viceversa, on which most of the calculus are based, are
written in Appendix \ref{sec-helements}.

\section{LRSUSY model, chargino sector, a brief description}
\label{sec-LRSUSY} The LRSUSY extension of the Standard Model
\cite{LRSUSY1,LRSUSY2,LRMODEL1,LRMODEL2,LRMODEL3,LRMODEL4,LRMODEL5,LRMODEL6,LRMODEL7},
based on the gauge group $SU(3)_L \times \SU(2)_L \times SU(2)_R
\times U_{{B-L}} \times P$ \cite{LRSUSYGG}, constitutes an
alternative to MSSM \cite{MSSM1}. In both supersymmetric models,
charginos are mixtures of charged gauginos and higgsinos fields.
However, the gauge sector of the LRSUSY model differs from gauge
sector of the MSSM in an extra neutral $Z^0_R$ and two charged
$W_R^\pm$ gauge bosons corresponding to the gauge group $SU(2)_R.$
Also, in the Higgs sector both models are different, in the LRSUSY
model the Higgs sector contains two bi-doublet fields associated to
the quark $u$ and $d:$
\begin{eqnarray} \label{eq:bi-doublet}\phi_{u,\,d}=\left(\begin{array}{cc}
\phi^0_{1}&\phi^+_{1}\nonumber \\
\phi^-_{2}&\phi^0_{2}
\end{array}\right)_{u,\,d}
\eeqa and four triplet fields, \beqa
\Delta_{L,\,R}=\left(\begin{array}{cc}
\frac{1}{\sqrt{2}}\,\Delta^{+}&\Delta^{++}\nonumber \\
\Delta^{0}&-\frac{1}{\sqrt{2}}\,\Delta^{+}
\end{array}\right)_{L,\,R},\eeqa
and \beqa \delta_{L,\,R}=\left(\begin{array}{cc}
\frac{1}{\sqrt{2}}\,\delta^{+}&\delta^{++}\nonumber \\
\delta^{0}&-\frac{1}{\sqrt{2}}\,\delta^{+}
\end{array}\right)_{L,\,R}.\eeqa
The Higgs $\phi_{u,d}$ transform as $(1/2,1/2,0),$ and  the Higgs
$\Delta_{L,\,R}$ transform as $(1,0,2)$ and $(0,1,2),$ respectively.
The triplet Higgs $\delta_{L,\,R}$ which transform as $(1,0,-2)$ and
$(0,1,-2),$ respectively, are introduced to cancel anomalies in the
fermionic sector.

The gauge group symmetry must be broken spontaneously in order to
generate masses to the quarks, leptons and gauge bosons and to break
parity. It can be achieved  by choosing the vacuum expectation
values (VEV's) of the Higgs fields in the form \cite{LRSUSY1} \be
\label{vevdelta}\langle \Delta_{L} \rangle = \langle
\delta_{L}\rangle = 0, \qquad \langle \Delta_{R} \rangle =
\begin{pmatrix}
0 & 0 \cr  \\ v_{\Delta_R} & 0 \cr \end{pmatrix}, \qquad \langle
\delta_{R} \rangle =
\begin{pmatrix}
0 & 0 \cr   \\ v_{\delta_R} & 0 \cr \end{pmatrix},  \ee

\be \label{vevkud} \langle \phi_{u} \rangle  =
\begin{pmatrix}
\kappa_{u}&0 \cr  \\
0&0 \cr \end{pmatrix}, \qquad \langle \phi_{d}
\rangle=\begin{pmatrix}
0&0 \cr   \\
0&\kappa_{d} \end{pmatrix}. \ee

There are  breakdowns at three different stages.  First parity is
breaking, no gauge bosons masses are generated. Next, the
spontaneous breaking of $SU(2)_{R}\times U(1)_{B-L}$ into $U(1)_Y,$
according to the VEV's  of the $\Delta_R $ and $\delta_R$ fields
given in Eq. \eqref{vevdelta}, generates masses for the gauge fields
$W^{\pm}_R, W^0_R$ and $V^0_R.$ Here, the two neutral states $W^0_R$
and $V^0_R $ mix yielding the physical field $Z^0_R$ and the
massless field $B^0.$  Then, the masses of the  weak bosons
$W^\pm_{L}$ and $W^0_L,$ as well as of $B_\nu,$ are generated at a
much lower energy scale by spontaneous breaking of $SU(2)_{L}\times
U(1)_{Y}$ into $U(1)_{\rm em},$ according the VEV's of $\phi_{u,d}$
given in Eq. \eqref{vevkud}. Once again, the neutral fields mix
forming the massless photon $A_\nu$ and the physical gauge field
$Z_L.$ Also at this stage,  the gaugino and Higgsino fields mix
forming the mass eigenstates of  neutralinos and charginos.

The chargino description is determined by the following  Lagrangian
involving the charged gaugino-higgsino fields: \be \mathcal{L}_{\rm
chargino} = - {1 \over 2}
\begin{pmatrix} {\psi^+}^T & {\psi^-}^T \cr
\end{pmatrix}
\begin{pmatrix} 0 & M^T \cr  M & 0\cr
\end{pmatrix}  \begin{pmatrix} \psi^+ \cr  \psi^- \cr
\end{pmatrix} + H.c,\ee with the chargino states
given by \be \label{psimas} \psi^+ = (\begin{matrix} -i \lambda^+_L
&  -i \lambda^+_R & {\tilde \phi}^+_{1u} &  {\tilde \phi}^+_{1d} &
{\tilde \Delta}^+_R \cr \end{matrix})^T \ee and  \be
\label{psimenos} \psi^- =(\begin{matrix} -i \lambda^-_L & -i
\lambda^-_R &  {\tilde \phi}^-_{2u} &  {\tilde \phi}^-_{2d} &
{\tilde \delta}^-_R \cr
\end{matrix})^T, \ee where  $\lambda^\pm_{L,R}$ are the   $SU(2)_{L,R}$
charged gauginos fields, ${\tilde \phi}^+_{1u},{\tilde \phi}^-_{2
u}$ and ${\tilde \phi}^+_{1 d},{\tilde \phi}^-_{2 d}$ are the
charged higgsino fields associated with the $u$ and $d$-quarks,
respectively. The charged right-handed higgsino fields are
represented by ${\tilde \Delta}^{+}_R$ and $ {\tilde \delta}^{-}_R.$
The chargino mass matrix $M$ is given by   \beqa M =
\begin{pmatrix}
M_{L} & 0 & 0 & \sqrt{2} {\tilde M}_L \cos\theta_\kappa & 0 \cr\nonumber \\
0 & M_{R} & 0 & \sqrt{2} {\tilde M}_R \cos\theta_\kappa & \sqrt{2}  M_1 \sin\theta_\upsilon \cr \nonumber \\
\sqrt{2} {\tilde M}_L \sin\theta_\kappa & \sqrt{2} {\tilde M}_R \sin\theta_\kappa & 0 & - \mu  & 0 \cr\nonumber \\
0 & 0 & - \mu & 0 & 0  \cr \nonumber \\
0 & \sqrt{2}  M_2 \cos\theta_\upsilon  & 0 & 0 & \mu_3 \cr
\end{pmatrix}, \label{eq:Mmatrix} \eeqa where we suppose  $M_L =
|M_L| e^{i \Phi_L},$ $\mu= |\mu| e^{i \Phi_{\mu}}, $  $\mu_3=
|\mu_3| e^{i \Phi_3}, $ \beqa
 {\tilde M}_L =  M_{W_L} e^{i {\tilde \Phi}_L}, \qquad
 {\tilde M}_R = {g_R \over g_L} M_{{W_L}}  e^{i {\tilde \Phi}_R}, \eeqa
 \beqa
 M_1 =  M_{{W_R}}  e^{i \Phi_1}, \qquad
 M_2 =  M_{{W_R}}  e^{i \Phi_2}, \eeqa where $M_{W_L}$
denotes  the mass of the left-handed gauge boson
  \be \label{eq:mass-WL} M_{W_L}= {g_L \over \sqrt{2}}
\sqrt{\kappa_u^2 + \kappa_d^2}\ee and  $M_{W_R}$ denotes the mass of
the right-handed gauge boson
  \be \label{eq:mass-WR} M_{W_R}= {g_R \over \sqrt{2}}
\sqrt{(v_{\Delta_R})^2 + (v_{\delta_R})^2}.\ee  Here $g_{L} $ and
$g_{R} $ are coupling constants associated to  the gauge groups
$SU(2)_{L}$ and $SU(2)_{R},$ respectively; $\mu$ and $\mu_3,$ are
fundamental Higgsino mass parameters, $M_L, M_R$  are fundamental
gaugino mass parameters.

The fundamental parameter  $\tan\theta_\kappa = k_u / k_d,$
represents the ratio between the vacuum expectation values of the
Higgs fields which couple to d- and u-type quark respectively. From
Eq. \eqref{eq:mass-WL} we deduce that  $\kappa_u$ and $\kappa_d$ can
be expressed in terms of $M_{W_L}, g_L$ and $\theta_\kappa$ in the
form \be \kappa_u = \sqrt{2} {M_{W_L} \over g_L} \sin\theta_\kappa,
\qquad  \kappa_d = \sqrt{2} {M_{W_L}\over g_L} \cos\theta_\kappa.
\ee  In the same way $\tan\theta_\upsilon = v_{\Delta_R} /
v_{\delta_R},$ represents the ratio between the vacuum expectation
values of the right-hand Higgs fields.   From Eq. \eqref{eq:mass-WR}
we deduce that $v_{\Delta_R}$ and $v_{\delta_R}$ can be expressed in
terms of $M_{W_R}, g_R$ and $\theta_\upsilon$ in the form

\be v_{\Delta_R} = \sqrt{2} {M_{W_R} \over g_R} \sin\theta_\upsilon,
\qquad v_{\delta_R} = \sqrt{2} {M_{W_R}\over g_R}
\cos\theta_\upsilon. \label{eq:vDeltaRvdeltaR}\ee

For the general  CP-violating case,  we are assuming that the
chargino mass matrix  is parameterized by thirteen real parameters,
namely, $ |M_L|,\Phi_L,|\mu|, \Phi_{\mu}, M_R ,{\tilde \Phi}_L,
{\tilde \Phi}_R, \Phi_1, \Phi_2, |\mu_3|, \Phi_3$
$\tan\theta_\kappa,$ and $\tan\theta_\upsilon.$


\section{Chargino  mass spectrum}
\label{sec-chargino-MASS} \setcounter{equation}{0} The physical
chargino mass eigenstates are related to the states given by the
Eqs. \eqref{psimas} and \eqref{psimenos} as \be \psi^{+}_{i}=
\sum_{j=1}^5  V_{ij} \,\chi^{+}_{j}, \qquad i=1,\ldots,5,  \ee \be
\, \psi^{-}_{i}= \sum_{j=1,5} U_{ij}\, \chi^{-}_{j} \qquad
i=1,\ldots,5. \ee The unitary matrices $U$ and $V$  satisfy \beqa
 M_{D} &=& U^T \,M \,V, \nonumber \\
&\equiv& \sum_{j=1}^{4}\, m_{{\tilde \chi}^{\pm}_{j}}\,E_{j},
\label{eq:VMV} \eeqa and
\be M^{2}_{D}  = V^{-1}\, M^{\dag} \,M\,V =  U^T \, M  \, M^\dag
\,U^\ast \equiv \sum_{j=1}^{5}\,m_{{\tilde
\chi}^{\pm}_{j}}^{2}\,E_{j}, \label{eq:MD2} \ee
where $(E_{j})_{5 \times 5 }$ are the basic matrices defined by \be
(E_{j})_{ik}= \delta_{ji} \, \delta_{jk}. \ee Here, we suppose that
the real eigenvalues of $M_{D}$ are ordered in the following way \be
\label{eq:orden-masas}  m_{{\tilde \chi}^{\pm}_{1}} <   m_{{\tilde
\chi}^{\pm}_{2}}  <  m_{{\tilde \chi}^{\pm}_{3}} <  m_{{\tilde
\chi}^{\pm}_{4}} < m_{{\tilde \chi}^{\pm}_{5}} .\ee

The chargino masses, at the tree level, are given by the positive
roots of the eigenvalues associated to either the Hermitian matrix $
H\equiv M^{\dag} \,M$ or the Hermitian matrix $\tilde H \equiv M \,
M^\dag.$ These eigenvalues can be obtained by solving the
characteristic equation associated to these matrices. In this
particular case, according to either  Eq. \eqref{eq:Hij} or Eq.
\eqref{eq:THij}, we can show that the characteristic equation can be
factorized in the form: \be ( m^2_{{\tilde \chi}^\pm_{j}}- |\mu|^2)
\, \left( (m^2_{{\tilde \chi}^\pm_{j}})^{4} - a \, (m^2_{{\tilde
\chi}^\pm_{j}})^{3} + b\, (m^2_{{\tilde \chi}^\pm_{j}})^{2} - c\,
m^2_{{\tilde \chi}^\pm_{j}} + d \right) = 0\, , \label{eq:CEQ} \ee
where \be a= {|\mu|}^2 + {|M_L|}^2 +
  2\,{|{\tilde M}_L|}^2 +
  2\,{|{\tilde M}_R |}^2 +
  {|\mu_3|}^2 +  2 M^2_{W_R}
   + M^2_R,  \label{eq:a-term}
\ee \beqa \nonumber  b &=& {|\mu |}^2\,{|\mu_3|}^2 +
  2\,{|{\tilde M}_R |}^2\,
   {|\mu_3 |}^2 +
  4\, \cos^2 (\theta_\kappa) \, {|{\tilde M}_L|}^4\,
   \sin^2 (\theta_\kappa) + 4 \, \cos^2 (\theta_\kappa) \,
   {|{\tilde M}_R|}^4\, \sin^2 (\theta_\kappa)  \nonumber \\ \nonumber &+&
  2\,\cos (\Theta_3 - \Theta_2 )\, |\mu| \, |M_L|\,
   {|{\tilde M}_L |}^2\,\sin (2\,{{\theta }_k}) +  2 \, \left( {|\mu|}^2  +
  2\, {|{\tilde M}_R|}^2  \sin^2 ({{\theta }_k} \right)  M^2_{W_R} \sin^2(\theta_\upsilon) \nonumber \\ \nonumber
   &+& 2 \left( {|\mu |}^2\, + 2\,{|{\tilde M}_R |}^2 \, \cos^2 ({{\theta }_k})\,
   \right) M^2_{W_R} \cos^2(\theta_\upsilon)  +  M^4_{W_R} \sin^2(2 \theta_\upsilon) \nonumber \\ \nonumber &+& 2\,
   \cos ( \Theta_2) \,
   |\mu | \, M_R \,{|{\tilde M}_R |}^2\,
   \sin (2\,{{\theta }_k})  -
  2\,\cos ( \Theta_1)\, |\mu_3 |  \, M_R  \, M^2_{W_R} \, \sin (2 \theta_\upsilon) \nonumber
\\ \nonumber & +&
  {|\mu|}^2\, M_R^2 +
  {| \mu_3 |}^2 \, M_R^2 +
  {|M_L |}^2 \, \left[ {|\mu |}^2 +
     2\, {|{\tilde M}_R |}^2 +
     {| \mu_3 |}^2 + 2 M^2_{W_R} + M_R^2 \right] \nonumber \\  &+&
  2 \,{| {\tilde M}_L |}^2\,
   \left[ {| \mu_3 |}^2 +
     {| {\tilde M}_R |}^2\,
      \sin^2 (2\,{{\theta }_k}) + 2 M^2_{W_R}
      + M_R^2 \right],  \label{eq:b-term} \eeqa
\beqa  \nonumber c&=& 2\, |\mu | \, {|{\tilde M}_R |}^2 \,
   | \mu_3 | \, \sin (2\,{{\theta }_k})
   \left[| \mu_3 | \, M_R \,
     \cos (\Theta_2 )\,
        -  M^2_{W_R} \, \cos( \Theta_1 - \Theta_2 ) \, \sin(2 \theta_\upsilon) \right]  \nonumber \\\nonumber &+&
  \left( |\mu|^2 +
     2\, {| {\tilde M}_L |}^2 \right) \,
   \left[ M^4_{W_R} \sin^2(2 \theta_\upsilon) -  2 \, | \mu_3 | \,
       M_R \, M^2_{W_R} \, \cos(\Theta_1) \, \sin(2 \theta_\upsilon) +
     {| \mu_3 |}^2 \,M_R^2 \right] \nonumber \\ \nonumber &+&
  {| M_L |}^2\,
   \biggl\{ {| {\tilde M}_R |}^4\,
      \sin^2 (2\,{{\theta }_k})  + 2
     {|\mu |}^2 \, M^2_{W_R}  +  M^4_{W_R} \, \sin^2(2 \theta_\upsilon) \nonumber
      \\ \nonumber &-&
     2 \, | \mu_3 | \, M_R \, M^2_{W_R} \,\cos ( \Theta_1 ) \, \sin (2 \theta_\upsilon) +
     {| \mu |}^2 \, M_R^2 \nonumber \\  \nonumber &+&
     2\, {| {\tilde M}_R |}^2\,
      \biggl[ { |\mu_3 |}^2 + 2 M^2_{W_R}
        \left( \cos^2 ( {{\theta }_k}) \, \cos^2( \theta_\upsilon) + \, \sin^2 ({{\theta }_k}) \, \sin^2( \theta_\upsilon) \right) + \cos
( \Theta_2 )\,
         |\mu | \, \sin (2\,{{\theta }_k})\, M_R
        \biggr] \nonumber \\  \nonumber &+& {| \mu_3 |}^2\,
      \left( {| \mu |}^2 + M_R^2 \right)  \biggr\}  +
  2\, |M_L | \, {|{\tilde M}_L |}^2\,
   \sin (2\,{{\theta }_k})\, \biggl[ \cos ( \phi_L -
        2\, {\tilde \phi}_ L +  2\, {\tilde \phi}_R) \,
      {| {\tilde M}_R |}^2 \, \sin (2\,{{\theta }_k})\,
      M_R \nonumber \\ &+& \nonumber  \cos ( \phi_ L -
        2\, {\tilde \phi}_L +  \phi_\mu ) \,
      |\mu | \, \left( {| \mu_3 |}^2 +  2 M^2_{W_R} +  M_R^2 \right)  \biggr] +
  {\sin (2\,{{\theta }_k})}^2\,  \nonumber \\ \nonumber &\times&
   \biggl[ {\left( {| {\tilde M}_L |}^2 +
           {| {\tilde M}_R |}^2 \right) }^2\,
      {| \mu_3 |}^2   +
     {| {\tilde M}_L |}^2\,
      \biggr( 2 {| {\tilde M}_R |}^2\, M^2_{W_R}
        +
        {| {\tilde M}_L |}^2\,  \left( 2 \, M^2_{W_R}  +
        M_R^2 \right)
          \biggl)  \biggr],  \label{eq:c-term} \eeqa
           \beqa
d&=&\left( |\mu |^2\, {| M_L |}^2 +
     4\, \cos^2 ({{\theta }_k}) \,
      {| {\tilde M}_L |}^4\, \sin^2 ({{\theta }_k})
     \right) \nonumber  \nonumber \\ \nonumber  &\times & \,\biggl[M^4_{W_R}  \sin^2(2 \theta_\upsilon) -
     2\, | \mu_3 | \, M_R \, M^2_{W_R}\, \cos ( \Theta_1 )\, \sin(2 \theta_\upsilon) +
     { |\mu_3 |}^2 \, M_R^2 \biggr] \nonumber \\ \nonumber &+&
  | M_L | \, {| {\tilde M}_R |}^2\,
   |\mu_3 |\, \sin^2 (2\,{{\theta }_k})\,
   \biggl[ | M_L | \, |{\tilde M}_R |^2\,
      | \mu_3 | +
     {| {\tilde M}_L |}^2\,
      \biggl(
        2\, | \mu_3 | \,
         M_R \, \cos ( \Theta_3 ) \,  \nonumber \\ \nonumber &-& 2\, M^2_{W_R} \, \cos( \Theta_1 - \Theta_3 )\,
          \, \sin(2 \theta_\upsilon)
          \biggr)  \biggr] +
  2\,  |\mu | \, | M_L | \,
   \sin ( 2\,{{\theta }_k}) \nonumber \\ \nonumber &\times& \biggl[ | M_L | \,
      {|{\tilde M}_R |}^2\,
      | \mu_3 | \,
      \biggl( | \mu_3 | \, M_R  \, \cos ( \Theta_2 )  - M^2_{W_R} \,  \cos  (
\Theta_1 - \Theta_2 )\,\sin(2 \theta_\upsilon) \biggr) \nonumber
\nonumber \\ &+&
     \cos ( \Theta_3 - \Theta_2 ) \,  {|{\tilde M}_L |}^2\,
      \biggl(M^4_{W_R}\, \sin^2(2 \theta_\upsilon)  -  2 \, | \mu_3 | \,
          M_R  M^2_{W_R}\, \cos ( \Theta_1 ) \,  \sin(2 \theta_\upsilon) +
        {| \mu_3 |}^2 \, M_R^2 \biggr)  \biggr]. \nonumber \nonumber \\  \label{eq:d-term}
           \eeqa
Here,  \beqa \label{Theta1} \Theta_1 &=& \Phi_1 + \Phi_2 - \Phi_3, \\
\Theta_2&=& 2 \, {\tilde \Phi}_R - \Phi_\mu \label{Theta2}\eeqa and
\be \Theta_3= \Phi_L - 2 \, {\tilde \Phi}_L + 2 {\tilde
\Phi}_R.\label{Theta3}\ee

Solving Eq. \eqref{eq:CEQ}, we get the  analytic expressions for the
chargino masses \beqa \label{eq:EIGU} m_{{\tilde \chi}^{\pm}_{1}}^2
, m_{{\tilde \chi}^{\pm}_{2}}^2 &=&
\frac{a}{4}-\frac{\varsigma}{2}\mp\frac{1}{2}\,\sqrt{\varrho -
\varpi-\frac{\lambda}{4\varsigma}},   \\
m_{{\tilde \chi}^{\pm}_{3}}^2&=& |\mu|^2 \label{eq:mass3} \\
m_{{\tilde \chi}^{\pm}_{4}}^2,m_{{\tilde \chi}^{\pm}_{5}}^2 &=&
\frac{a}{4}+\frac{\varsigma}{2}\mp\frac{1}{2}\,\sqrt{\varrho
-\varpi+\frac{\lambda}{4\varsigma}},\label{eq:EIGV} \eeqa
where \beqa \varsigma
&=& \nonumber \sqrt{{\varrho \over 2} + \varpi},\nonumber \\
\nonumber \varpi &=&\frac{\epsilon}{3 \ 2^{\frac{1}{3}}} +
\frac{(2^{\frac{1}{3}}\,\gamma)}{3\,\epsilon},\nonumber \\ \nonumber
\epsilon &=& (\delta+ \sqrt{\delta^2 - 4 \gamma^3})^{\frac{1}{3}},
\nonumber \\ \nonumber \varrho &=&
\frac{a^{2}}{2}-\frac{4b}{3},\nonumber \\\nonumber
 \lambda &=& \nonumber a^{3}- 4\,a\,b + 8\,c\nonumber \\ \nonumber
\gamma &=& \nonumber b^{2}- 3\,a\,c+12\,d, \nonumber \\ \delta &=& \
2\,b^{3}- 9\,a\,b\,c+27\,c^{2}+ 27\,a^{2}\,d-72\,b\,d.
\label{eq:relation-parameters}\eeqa

According to the characteristic equation \eqref{eq:CEQ}, it is
always possible to find a neighborhood in the fundamental parameter
space where one of the chargino masses takes the value $|\mu|.$
Assuming the neighborhood of the fundamental parameter space for
$m_{{\tilde \chi}^{\pm}_{3}}=|\mu|,$ the physical masses given in
Eqs. (\ref{eq:EIGU}-\ref{eq:EIGV}) are automatically arranged
according to their magnitude, from the lightest to the heaviest. In
regions where a new chargino mass different from $m_{{\tilde
\chi}^{\pm}_{3}}$ takes the value $|\mu|,$ we must simply redefine
the masses' suffixes given in Eqs. \eqref{eq:EIGU} and
\eqref{eq:EIGV}, without altering the increasing order, taking into
account the suffix of this new chargino mass.

Notice that, the expressions for the chargino masses given in Eqs.
(\ref{eq:EIGU}-\ref{eq:EIGV}) include the impact of the given
CP-violating phases. The chargino masses only depend on the phase
combinations (\ref{Theta1}-\ref{Theta3}) which  describe the
influence of the charged right-handed higgisno fields upon the
chargino masses. Thus, they constitutes a generalization of some
results found in the literature
\cite{nalvar8,Frank-neu-char-masses-1,Frank-neu-char-masses-2}

\section{Diagonalizing matrices \boldmath $V$  and $U^\ast$}
\label{sec-diagonalizing} From Eq. \eqref{eq:MD2}, we can show that
the entries of the diagonalizing matrix  $V,$ for a fixed $\ell$
($\ell= 1, 2,\ldots,4$ or $5$), are given by \be
\label{eq:GEN-EVEVij} V_{ij} = {\Delta^{(\ell)}_{ij} \over
\Delta^{(\ell)}_{\ell j}} \, {| \Delta^{(\ell)}_{\ell j} | \, e^{- i
\vartheta_{\ell j}} \over \sqrt{ \sum_{k=1}^5 | \Delta^{(\ell)}_{kj}
|^2}},  \qquad i,j=1,\ldots,5,\ee  where \be
\label{eq:GEN-deltaellj} \Delta^{(\ell )}_{\ell j}(H) = {\rm Det}
\left(H^{(\ell,\ell)}  - m_{{\tilde \chi}^{\pm}_{j}}^2 \, I_4
\right) \ee and $\Delta^{(\ell )}_{ij} (H), \, i=1, \ldots,5,$ $i\ne
\ell,$ are formed from $\Delta^{(\ell)}_{\ell j}(H)$ by substituting
the $1,\ldots,4$-th columns by the $4 \times 1$ matrices obtained
from $ \biggl(\begin{smallmatrix} - H_{1 \ell} \cr  \vdots \cr  -
H_{5 \ell}\cr
\end{smallmatrix}\biggr) $ by eliminating the $\ell$-th row,
respectively.  $H^{(\ell ,\ell)}$ is the minor matrix formed from
$H$ if we eliminate the $\ell$-th row and the $\ell$-th column.
$I_4$ is the $4 \times 4$ identity matrix.

In the same way we can write  the matrix elements of $U^\ast,$ we
get \be \label{eq:GEN-EVEUij} U^\ast_{ij} = {{\tilde
\Delta}^{(\ell)}_{ij} \over {\tilde \Delta}^{(\ell)}_{\ell j}} \, {|
{\tilde \Delta}^{(\ell)}_{\ell j} | \, e^{ i {\tilde
\vartheta}_{\ell j}} \over \sqrt{ \sum_{k=1}^4 | {\tilde
\Delta}^{(\ell)}_{kj} |^2}}, \qquad i,j=1,\ldots,5,\ee where
${\tilde \Delta}^{(\ell)}_{ij} (\tilde H )\equiv
\Delta^{(\ell)}_{ij}(\tilde H ).$

Thus, to know the $V$-type and $U^\ast$-type diagonalizing matrices
we only need to compute the $\Delta^{(\ell)}_{\alpha j}$'s and
${\tilde \Delta}^{(\ell)}_{\alpha j}$'s basic quantities,
respectively. From the definition for these quantities, assuming
that $m_{{\tilde \chi}^{\pm}_{3}}=|\mu|,$  and taking $\ell=1,$ we
get ($j=1,2,4,5$) \beqa \Delta_{1j} &=& \left(m^2_{{\tilde
\chi}^\pm_j } - {|\mu |}^2 \right) \,
  \biggl\{m^6_{{\tilde \chi}^\pm_j} -  m^4_{{\tilde \chi}^\pm_j } \, \biggl[ M^2_R +
       2\,  \cos^2 (\theta_\kappa) \, { |{\tilde M}_L |}^2 + 2\,{ | {\tilde M}_R |}^2
       +  g^2_R \, \left( (v_{\Delta_R})^2 + (v_{\delta_R})^2 \right) \nonumber \\ &+&
       {|\mu |}^2 + {| \mu_3 |}^2 \biggr]  +
    m^2_{{\tilde \chi}^\pm_j }\, \biggl[ g^4_R \, (v_{\Delta_R})^2\,(v_{\delta_R})^2 +
       M^2_R\, {| \mu |}^2  +
       g^2_R \, \left( (v_{\Delta_R})^2 + (v_{\delta_R})^2 \right) \,
        {| \mu |}^2  - 2\, g^2_R \, M_R \, \nonumber \\ &\times&
        \cos ( \Theta_1 ) \, v_{\Delta_R}\,
        v_{\delta_R}\, | \mu_3 | +
       \left( M^2_R + {| \mu |}^2 \right) \, {| \mu_3 |}^2 +
       2\, \cos^2 (\theta_\kappa) \, {| {\tilde M}_L |}^2\,
        \bigl( M^2_R + g^2_R \, \nonumber \\ &\times&
           \left( (v_{\Delta_R})^2 + (v_{\delta_R})^2 \right) +
          {| \mu_3 |}^2 + 2\,{| {\tilde M}_R |}^2\, \sin^2 (\theta_\kappa)
          \bigr)  + {| {\tilde M}_R |}^4\, \sin^2 (2\,\theta_\kappa)  +
       2\, {| {\tilde M}_R |}^2\, \bigl( {| \mu_3 |}^2 +
          g^2_R \, \nonumber \\ &\times&\left( \cos^2 (\theta_\kappa) \, (v_{\delta_R})^2 +
             (v_{\Delta_R})^2\, \sin^2 (\theta_\kappa) \right)  +
          M_R\,\cos ( \Theta_2 ) \, | \mu |\,
           \sin (2\,\theta_\kappa) \bigr)  \biggr]  -
    {| \mu |}^2\, \nonumber \\ &\times& \left( g^4_R \, | v_{\Delta_R}|^2\,
        (v_{\delta_R})^2 - 2\, g^2_R \,M_R\,
        \cos ( \Theta_1 ) \,  v_{\Delta_R}\,
         v_{\delta_R} \, | \mu_3 | +
       M_R^2 \, {| \mu_3 |}^2 \right) -
    2\, \cos^2 (\theta_\kappa) \nonumber \\ &\times&  {| {\tilde M}_L |}^2\,
     \biggl( g^4_R \, (v_{\Delta_R})^2\,(v_{\delta_R})^2 -
       2\, g^2_R \,M_R \, \cos ( \Theta_1 ) \,
        v_{\Delta_R}\, v_{\delta_R} \, |\mu_3|  +
       M^2_R\, {|\mu_3|}^2 +
       2\,{| {\tilde M}_R |}^2  \nonumber \\ &\times&\left( {g_R}^2\,(v_{\Delta_R})^2 +
          {| \mu_3 |}^2 \right) \, \sin^2 (\theta_\kappa) \biggr) - {| {\tilde M}_R |}^4\,
     {| \mu_3 |}^2\, \sin^2 (2\,\theta_\kappa) +
    2\,{| {\tilde M}_R |}^2\,| \mu | \,| \mu_3 |\, \,\sin
    (2\,\theta_\kappa) \nonumber \\&\times&
    \, \left( {g_R}^2\,\cos (\Theta_1 - \Theta_2 )\, v_{\Delta_R} \, v_{\delta_R}
       - M_R\,\cos ( \Theta_2 ) \, | \mu_3 |
       \right)   \biggr\}, \label{eq:delta1j}\eeqa

 \beqa \Delta_{2j}&=& 2\,|{\tilde M}_L |\,|{\tilde M}_R |\, e^{-
i \,\left( \Phi_1 + \Phi_2 + {\tilde \Phi}_L + {\tilde \Phi}_R +
        \Phi_\mu \right) } \,
    \left( m^2_{{\tilde \chi}^\pm_j} - {|\mu |}^2 \right) \,
    \biggl\{ e^{i \,\left( \Phi_1 + \Phi_2 + 2\,{\tilde \Phi}_L + \Phi_\mu \right)
    }\,
       m^4_{{\tilde \chi}^\pm_j}\,\sin^2 (\theta_\kappa) \nonumber \\ &-& e^{i \,\left( \Phi_1 + \Phi_2 \right) }\,
       m^2_{{\tilde \chi}^\pm_j}\,\biggl[ e^{i \,\left( \Phi_L + \Phi_\mu \right) }\,\cos (\theta_\kappa)\,
          |M_L |\,  \left( e^{i \,\Phi_\mu}\,|\mu |\,\sin (\theta_\kappa)  -  e^{ 2\,i  \,{\tilde \Phi}_R}\,M_R\,
               \cos (\theta_\kappa) \right)
            \nonumber \\ &+&
         e^{ 2\,i \,{\tilde \Phi}_L}\,\sin (\theta_\kappa)\,
          \biggl( e^{ 2\,i \,{\tilde \Phi}_R}\,M_R\,\cos (\theta_\kappa)\,
             |\mu | +  e^{i \,\Phi_\mu}\, \sin (\theta_\kappa)\,
             \bigl( 2\,{\cos (\theta_\kappa)}^2\,{|{\tilde M}_L |}^2 +
               2\,\cos^2 (\theta_\kappa) \, \nonumber \\ & \times & \, {|{\tilde M}_R |}^2 +
               g^2_R\,(v_{\Delta_R})^2 + {|\mu_3 |}^2 \bigr) \,
              \biggr)  \biggr] +
      \cos (\theta_\kappa)\,\biggl[ e^{i \,\left( \Phi_L + \Phi_\mu \right) }\,
          |M_L |\,\biggl( e^{i \,\left( \Phi_3 + 2\,{\tilde \Phi}_R \right) }\,
             g^2_R\,\cos (\theta_\kappa)\, \nonumber \\ &\times& v_{\Delta_R}\,
             v_{\delta_R}\,|\mu_3 | -
            e^{i \,\left( \Phi_1 + \Phi_2 + 2\,{\tilde \Phi}_R \right) }\,M_R\,
             \cos (\theta_\kappa)\,{|\mu_3 |}^2 +
            e^{i \,\left( \Phi_1 + \Phi_2 + \Phi_\mu \right) }\,g^2_R\,
             (v_{\Delta_R})^2\,|\mu |\,\sin (\theta_\kappa) \nonumber \\ &+&
            e^{i \,\left( \Phi_1 + \Phi_2 + \Phi_\mu \right) }\,|\mu |\,
             {|\mu_3 |}^2\,\sin (\theta_\kappa) \biggr)  \nonumber +
         e^{ 2\,i  \,{\tilde \Phi}_L}\,\sin (\theta_\kappa)\,
          \biggl(
            e^{i \,\left( \Phi_1 + \Phi_2 + \Phi_\mu \right) }\,g^2_R\,
             {|{\tilde M}_L |}^2\,(v_{\Delta_R})^2\,\sin (2\,\theta_\kappa) \nonumber \\&-&  e^{i \,\left( \Phi_3 + 2\,{\tilde \Phi}_R \right) }\,g^2_R\,
               v_{\Delta_R}\,v_{\delta_R}\,|\mu |\,
               |\mu_3 |   +
            e^{i \,\left( \Phi_1 + \Phi_2 \right) }\,{|\mu_3 |}^2\,
             \bigl[ e^{ 2\,i  \,{\tilde \Phi}_R}\,M_R\,|\mu | +
               e^{i \,\Phi_\mu}\, \sin (2\,\theta_\kappa) \, \nonumber \\ &\times& \left( {|{\tilde M}_L |}^2 +
                  {|{\tilde M}_R |}^2 \right)  \bigr]  \biggr)
                  \biggr]
      \biggr\}, \label{eq:delta2j}  \eeqa

      \beqa
\Delta_{3j}&=&0, \label{eq:delta3j}
      \eeqa
\beqa \Delta_{4j}&=& -   e^{- i \,\left( {\tilde \Phi}_L +
2\,{\tilde \Phi}_R + \Phi_\mu \right) } {\sqrt{2}}\,|{\tilde M}_L |
\,  \left(m^2_{{\tilde \chi}^\pm_j} - {|\mu |}^2 \right) \,
  \biggl\{
       \left( m^2_{{\tilde \chi}^\pm_j} - g^2_R\,(v_{\Delta_R})^2 - {|\mu_3 |}^2
         \right) \nonumber \\ &\times& \biggl[ e^{i \,\left( \Phi_L + 2\,{\tilde \Phi}_R + \Phi_\mu \right) }\,
          \cos (\theta_\kappa)\,|M_L |\,
          \left( M^2_R - m^2_{{\tilde \chi}^\pm_j} + g^2_R\,(v_{\delta_R})^2 +
            2\,{|{\tilde M}_R |}^2\,{\sin (\theta_\kappa)}^2 \right) \nonumber \\
            &-&
         e^{2\,i \,{\tilde \Phi}_L}\,\sin (\theta_\kappa)\,
          \left( e^{ 2\,i  \,{\tilde \Phi}_R}\,
             \left( M^2_R  - m^2_{{\tilde \chi}^\pm_j} + g^2_R\,(v_{\delta_R})^2 \right) \,
             |\mu | + e^{i \,\Phi_\mu}\,M_R\,{|{\tilde M}_R |}^2\,
             \sin (2\,\theta_\kappa) \right)  \biggr]   \nonumber \\
       &+& e^{- i \,\left( \Phi_1 + \Phi_2 + \Phi_3  \right) }
          \, g^2_R \, \left( e^{i \,\Phi_3}\,M_R\,v_{\Delta_R} +
         e^{i \,\left( \Phi_1 + \Phi_2 \right) }\,v_{\delta_R}\,
          |\mu_3 | \right) \,
       \biggl[ e^{i \,\left( \Phi_L + 2\,{\tilde \Phi}_R + \Phi_\mu \right) }\,
          \cos (\theta_\kappa)\, \nonumber \\  &\times& |M_L |\,
          \left( e^{i \,\left( \Phi_1 + \Phi_2 \right) }\,M_R\,
             v_{\Delta_R} + e^{i \,\Phi_3}\,v_{\delta_R}\,
             |\mu_3 | \right) -
         e^{ 2\,i  \,{\tilde \Phi}_L}\,\sin (\theta_\kappa)\,
          \biggl( e^{i \,\left( \Phi_1 + \Phi_2 + 2\,{\tilde \Phi}_R \right) } \,M_R\,
             \nonumber \\ &\times& v_{\Delta_R}\,|\mu | +
            e^{i \,\left( \Phi_3 + 2\,{\tilde \Phi}_R \right) }\,v_{\delta_R}\,
             |\mu |\,|\mu_3 | +
            e^{i \,\left( \Phi_1 + \Phi_2 + \Phi_\mu \right) }\,
             {|{\tilde M}_R |}^2\,v_{\Delta_R}\,\sin (2\,\theta_\kappa) \biggr)
         \biggr] \biggr\}, \label{eq:delta4j} \eeqa

         \beqa
\Delta_{5j}&=&  2 \,g_R\,|{\tilde M}_L|\,|{\tilde M}_R|\,
    \left( m^2_{{\tilde \chi}^\pm_j} - {|\mu |}^2 \right) \,e^{- i \,\left( \Phi_1 + \Phi_3 + {\tilde \Phi}_L +
{\tilde \Phi}_R +
        \Phi_\mu \right) }  \Biggl\{  m^2_{{\tilde \chi}^\pm_j} \Biggl[ e^
             {i \,\left( \Phi_3 + \Phi_L + 2\,{\tilde \Phi}_R + \Phi_\mu \right) }  \nonumber \\ &\times&
            {\cos (\theta_\kappa)}^2\,|M_L |\,v_{\Delta_R} +
           e^{\left( 2\,i  \right) \,{\tilde \Phi}_L}\,\sin
           (\theta_\kappa)\,
            \biggl(
              e^{i \,\left( \Phi_3 + \Phi_\mu \right) }\,M_R\,
               v_{\Delta_R}\,\sin (\theta_\kappa)  +
              e^{i \,\left( \Phi_1 + \Phi_2 + \Phi_\mu \right) } \nonumber \\
              &\times&
               v_{\delta_R}\,|\mu_3 |\,\sin (\theta_\kappa) - e^{i \,\left( \Phi_3 + 2\,{\tilde \Phi}_R \right) }\,\cos (\theta_\kappa)\,
                 v_{\Delta_R}\,|\mu |   \biggr)
           \Biggr] -  \cos (\theta_\kappa) \,
       \Biggl[ e^{ 2\,i  \,{\tilde \Phi}_L} \, \sin (\theta_\kappa)\,  \nonumber \\ &\times&
          \biggl[ e^{i \,\left( \Phi_1 + \Phi_2 \right) }\,v_{\delta_R}\,
             |\mu_3 |\,\left( e^{ 2\,i  \,{\tilde \Phi}_R}\,M_R\,
                |\mu | + e^{i \,\Phi_\mu}\,
                \left( {|{\tilde M}_L|}^2 + {|{\tilde M}_R|}^2 \right) \,
                \sin (2\,\theta_\kappa) \right) -  e^{i \,\Phi_3}\,v_{\Delta_R}\, \nonumber \\ &\times&
               \left( e^{2\,i \,{\tilde \Phi}_R}\,g^2_R\,
                  (v_{\delta_R})^2\,|\mu | -
                 e^{i \,\Phi_\mu}\,M_R\,{|{\tilde M}_L|}^2\,
                  \sin (2\,\theta_\kappa) \right) \biggr]  +
         e^{i \,\left( \Phi_L + \Phi_\mu \right) }\,|M_L |\, \biggl[ e^{i \,\Phi_3}\,v_{\Delta_R} \nonumber \\ &\times&
             \left( e^{ 2\,i  \,{\tilde \Phi}_R}\,g^2_R\,\cos (\theta_\kappa)\,
                (v_{\delta_R})^2 +
               e^{i \,\Phi_\mu}\,M_R\,|\mu |\,\sin (\theta_\kappa) +
               e^{ 2\,i \,{\tilde \Phi}_R}\,{|{\tilde M}_R|}^2\,
                \sin (\theta_\kappa)\,\sin (2\,\theta_\kappa) \right)
                \nonumber \\ &-&  e^{i \,\left( \Phi_1 + \Phi_2 \right) }\,v_{\delta_R}\,
               |\mu_3 |\,
               \left( e^{2\,i \,{\tilde \Phi}_R}\,M_R\,\cos (\theta_\kappa) -
                 e^{i \,\Phi_\mu}\,|\mu |\,\sin (\theta_\kappa) \right) \biggr]  \Biggr]\Biggr\},   \label{eq:delta5j} \eeqa
                 where $v_{\Delta_R}$
                 and $v_{\delta_R}$ are given by Eq.
                 \eqref{eq:vDeltaRvdeltaR}.
                 The expressions for ${\tilde \Delta}_{kj}, \,
                 k=1,\ldots,5, $ are obtained  by
                 interchanging $v_{\Delta_R} \leftrightarrow v_{\delta_R},$  $\Phi_1 \leftrightarrow
                 \Phi_2$ and $\sin \theta_\kappa \leftrightarrow \cos
                 \theta_\kappa$ in the equations
                  $\Delta^\ast_{1j}, \Delta^\ast_{2j}, \Delta^\ast_{4j},
                 \Delta^\ast_{3j}$  and $\Delta^\ast_{5j}$ given above,  respectively.

The remaining  $ \Delta_{ij}$ and ${\tilde \Delta}_{ij}$ factors can
be deduced taking into account the properties of $V$ and $U^\ast.$
Equation \eqref{eq:delta3j} implies that all the entries of the
third row of the  $V$  matrix, except $V_{33},$ are zero. Since $V$
is an invertible matrix,  $V_{33}$ must be different from zero. In
addition, $V$ is unitary, then all the entries of the third column
of this matrix, except $V_{33},$  must be equal to zero.
consequently the norm of $V_{33}$ is equal to 1. From Eq.
\eqref{eq:GEN-EVEVij} we deduce $\Delta_{\alpha 3}=0,
\alpha=1,2,4,5.$  Similarly, using the same arguments,  for the
matrix $U^\ast,$ we have $U^\ast_{\alpha 3}=0, \alpha=1,2,3,5$ and
$U^\ast_{4 \beta}=0, \beta=1,2,4,5.$ Also, the norm of $U^\ast_{43}$
must be equal to 1. Finally, from Eq. \eqref{eq:GEN-EVEUij} we
deduce ${\tilde \Delta}_{4 \beta}=0, \beta=1,2,4,5.$

\section{Generalized projector formalism}
\label{sec-projector-formalism}
 Based on the diagonalization process  of the chargino mass matrix studied in the previous section,
 we implement a projector method which can be easily generalized to deal with any symmetric or non-symmetric
 mass matrix. This method provides both a system of basic equations connecting the fundamental parameters with the
physical chargino masses,  the reduced projectors, and the
eigenphases and a set of equivalences (when we combine with the
Jarlskog's projector formulation). These equations and equivalences,
as we will see in the next sections, are useful for both to
disentangle some relevant fundamental parameters from the
 rest of the parameters (see
Section \ref{sec-det-parameters}) and to obtain a systematic
inversion process based on the measurement of some suitable set of
physical observables. (see Section \ref{sec-par-inversion}).

\subsection{Reduced projectors}
\label{sec-projector-formalism-uno}Defining the type-$V$ reduced as
projectors \be p^{(\ell)}_{j \alpha} = {{(\Delta^{(\ell)}_{\alpha
j})}^\ast \over {(\Delta^{(\ell)}_{\ell j})}^\ast}
\label{eq:def-red-proj-v} \ee  we can express the entries of the
diagonalizing matrix $V$ given in Eq. \eqref{eq:GEN-EVEVij} in the
compact form  \be \label{eq:Vaj} V_{\alpha j} = \sqrt{P^V_{j\ell
\ell} \over \eta^{(\ell)}_j} (p^{(\ell)}_{j \alpha})^\ast, \ee where
$\eta^{(\ell)}_j \equiv e^{2 i \vartheta^{(\ell)}_j }$ stands for
the type-$V$  eigenphases and \be \label{eq:CPjalpha-V} P^V_{j
\alpha \beta} =  P^V_{j \ell \ell} \, \, (p^{(\ell)}_{j
\alpha})^\ast  \, p^{(\ell)}_{j \beta} = { (p^{(\ell)}_{j
\alpha})^\ast \, p^{(\ell)}_{j \beta} \over \sum_{k=1}^5
|p^{(\ell)}_{jk}|^2 },\ee stands for the type-$V$ projectors (note
that in general $p^{(\ell)}_{j  \ell }= 1$ whatever $
\Delta^{(\ell)}_{\ell j} \ne 0$).

Similarly, by defining the type-$U^\ast$ reduced projectors \be
{\tilde p}^{(\ell)}_{j \alpha} = {{({\tilde \Delta}^{(\ell)}_{\alpha
j})}^\ast \over {({\tilde \Delta}^{(\ell)}_{\ell j})}^\ast},
\label{eq:def-red-proj-u} \ee  we can express the entries of the
diagonalizing matrix $U^\ast$ given in Eq. \eqref{eq:GEN-EVEUij} in
the compact form \be \label{eq:Uaj} U^\ast_{\alpha j} =
\sqrt{P^{U^\ast}_{j \ell \ell } \over {\tilde \eta}^{(\ell)}_j}
({\tilde p}^{(\ell)}_{j \alpha})^\ast, \ee where ${\tilde
\eta}^{(\ell)}_j \equiv e^{- 2 i {\tilde \vartheta}^{(\ell)}_j }$
stands for the type-$U^\ast$ CP eigenphases and \be
\label{eq:CPjalpha-U} P^{U^\ast}_{j \alpha \beta} =  P^{U^\ast}_{j
\ell \ell} \, \, ({\tilde p}^{(\ell)}_{j \alpha})^\ast \, {\tilde
p}^{(\ell)}_{j \beta} ={ ({\tilde p}^{(\ell)}_{j \alpha})^\ast \,
{\tilde p}^{(\ell)}_{j \beta} \over \sum_{k=1}^5  |{\tilde
p}^{(\ell)}_{jk}|^2},\ee stands for the type-$U^\ast$ projectors.

Note that the reduced projector are not all independent. Since $V$
and $U^\ast$ are unitary matrices, from \eqref{eq:Vaj} and
\eqref{eq:Uaj}, we get the  constraints \be  P^V_{i \ell \ell} \,
\sum_{k=1}^{5} p^{(\ell)}_{ik} \, (p^{(\ell)}_{jk})^\ast =
\delta_{ij} \label{eq:sysV} \ee and \be
 P^{U^\ast}_{i \ell \ell} \,  \sum_{k=1}^{5} {\tilde p}^{(\ell)}_{ik} ({\tilde
p}^{(\ell)}_{jk})^\ast = \delta_{ij}. \label{eq:tsysU}\ee For $i=j,$
Eqs. \eqref{eq:sysV} and \eqref{eq:tsysU} verify identically. For
$i,j=1,\ldots,5, \mid \, \, i>j,$  each one represent a system of
ten complex algebraic equations which can be used to reduce up to
$20$ the number of real independent parameters on each set of
reduced projectors.

On the other hand, from the particular  structure of the $H$ and
${\tilde H}$ matrices, we see that the entries of the $V$ and
$U^\ast$ matrices are related as follows \be \label{eq:UIJ}
U^\ast_{ij}= {1 \over m_{{\tilde \chi}^{\pm}_{j}} } \sum_{k=1}^5
M_{i k} V_{kj}\ee and \be \label{eq:VIJ} V_{ij}= {1 \over m_{{\tilde
\chi}^{\pm}_{j}} } \sum_{k=1}^5 M^\dagger_{i k} U^\ast_{kj}. \ee
Hence, inserting Eqs. \eqref{eq:Vaj} and \eqref{eq:Uaj}  into Eq.
\eqref{eq:UIJ} and Eq. \eqref{eq:VIJ} we deduce the fundamental
basic equations  ($\alpha =1,\ldots,5$) \be
\label{eq:gen-inversionU} m_{{\tilde \chi}^{\pm}_{j}} \,
\zeta^{(\ell)}_j \, \sqrt{P^{U^\ast}_{j \ell \ell}\over P^V_{j \ell
\ell}} \, ({\tilde p}^{(\ell)}_{j \alpha})^\ast = \sum_{\beta=1}^5
M_{\alpha \beta} \, (p^{(\ell)}_{j \beta})^\ast \ee and \be
\label{eq:gen-inversionV} m_{{\tilde \chi}^{\pm}_{j}} \,
\zeta^{(\ell)}_j \, \sqrt{ P^V_{j \ell \ell} \over P^{U^\ast}_{j
\ell \ell}} \, p^{(\ell)}_{j \alpha} = \sum_{\beta=1}^5 M_{\beta
\alpha} \, {\tilde p}^{(\ell)}_{j \beta}, \ee respectively.  Here,
$\zeta^{(\ell)}_j \equiv \sqrt{{\eta^{(\ell)}_j \over {\tilde
\eta}^{(\ell)}_j}}$ stands for the global eigenphases.

Equations \eqref{eq:gen-inversionU} and \eqref{eq:gen-inversionV}
represent, for fixed $j,$ a system of ten complex algebraic
equations serving to determine the fundamental parameters of the
model in terms of the reduced projectors, the chargino physical
masses $m_{{\tilde \chi}^\pm_j }$ and the eigenphases, and vice
versa.

Note that, in Eqs. \eqref{eq:Vaj} and \eqref{eq:Uaj} as well as in
Eqs. \eqref{eq:gen-inversionU} and \eqref{eq:gen-inversionV},
without any loss of generality,  we could choose  the $U^\ast$-type
eigenphases either ${\tilde \eta}^{(\ell)}_j=1, j=1,\ldots,5,$ such
that $\zeta^{(\ell)}_j= \sqrt{\eta^{(\ell)}_j}$;  $({\tilde
\eta}^{(\ell)}_j)^\ast = \eta^{(\ell)}_j, j=1,2,\ldots,5,$ such
$\zeta^{(\ell)}_j=\eta^{(\ell)}_j,$ or any other suitable choice
allowing us to eliminate five superfluous parameters.

Note that the generalization of the projector formalism to any
number of charginos  is direct, for instance, for  $n$ charginos,
$n=2,\ldots,$ we only need to consider $\ell$ a fixed number between
$1$ and $n,$ $I_{n-1}$ in place of $I_4$ and the subindex $\alpha,
\beta, j$ running from  $1$ to $n.$

\subsection{Reduced projectors in terms of the fundamental parameters
and eigenphases} From Eqs. \eqref{eq:gen-inversionU} and
\eqref{eq:gen-inversionV}, we can express the reduced projectors in
terms of the fundamental parameters and eigenphases without any
dependence on the parameters $|M_L|$ and $\Phi_L.$ Indeed, choosing
$\ell=1,$ and  combining the corresponding equations for
$j=1,2,4,5,$ and $\alpha=2,3,4,5,$ we get the non trivial reduced
projectors: \beqa p_{j2}&=& { 2 \,e^{- i \,\left( \Phi_3 + \Phi_\mu
\right) } {\tilde M}^\ast_L \, {\tilde M}^\ast_R \over {\tilde {\cal
D}}_j } \biggl\{ e^{ 2\,i \,{\tilde
\Phi}_L}\,{\sqrt{\frac{P^{U^\ast}_{j11}}{P^{V}_{j11}}}}\,\xi^\ast_j\,
   \cos (\theta_\kappa)\,m_{{\tilde \chi}^\pm_j}\,\biggl[ e^{i \,\left( \Phi_3 + 2\,{\tilde \Phi}_R \right) }\,|\mu
        |\, \,\sin (\theta_\kappa) \nonumber \\ &\times &
      \left(  g^2_R\,(v_{\Delta_R})^2 + {|\mu_3 |}^2 - m^2_{{\tilde \chi}^\pm_j}
        \right) + e^{i \,\Phi_\mu}\,\cos
   (\theta_\kappa)\,  \biggl( e^{i \,\left( \Phi_1 + \Phi_2 \right) }\,g^2_R\,
         v_{\Delta_R}\,v_{\delta_R}\,|\mu_3 | \nonumber \\&+&
        e^{i \,\Phi_3}\,M_R\,\left( m^2_{{\tilde \chi}^\pm_j} - {|\mu_3 |}^2 \right)
        \biggr)   \biggr]  -
  e^{i \,\Phi_\mu}\,\sin (\theta_\kappa)\,
   \biggl[e^{i \,\Phi_\mu}\,\cos (\theta_\kappa)\,|\mu |\,
        \biggl(
          e^{i \,\Phi_3}\,M_R\, \nonumber \\ &\times& \left( m^2_{{\tilde \chi}^\pm_j} - {|\mu_3 |}^2 \right) + e^{i \,\left( \Phi_1 + \Phi_2 \right) }\,g^2_R\,
           v_{\Delta_R}\,v_{\delta_R}\,|\mu_3 |
          \biggr) -  e^{i \,\left( \Phi_3 + 2\,{\tilde \Phi}_R \right)
          } \nonumber \\ &\times &
      \biggl( {m_{{\tilde \chi}^\pm_j}}^4\,\sin (\theta_\kappa) -
        m^2_{{\tilde \chi}^\pm_j}\,\left( 2\,{\cos (\theta_\kappa)}^2\,{|{\tilde M}_R |}^2 +
           g^2_R\,(v_{\Delta_R})^2 + {|\mu_3 |}^2 \right) \,
         \sin (\theta_\kappa) \nonumber \\ &+& \cos (\theta_\kappa)\,{|{\tilde M}_R |}^2\,
         {|\mu_3 |}^2\,\sin (2\,\theta_\kappa) \biggr)  \biggr]\biggr\}\label{eq:pj2}\eeqa

         \beqa
p_{j4} &=& { \sqrt{2} {\tilde M}^\ast_L \,  \over {\tilde {\cal
D}}_j }\, \biggl\{ e^{ 2\,i \,{\tilde \Phi}_L}
\,{\sqrt{\frac{P^{U^\ast}_{j11}}{P^{V}_{j11}}}}\,\xi^\ast_j\,
   \cos (\theta_\kappa)\,m_{{\tilde \chi}^\pm_j} \,\biggl[ m^4_{{\tilde \chi}^\pm_j} +
     g^4_R\,(v_{\Delta_R})^2\,(v_{\delta_R})^2  +
     M^2_R\,{|\mu_3 |}^2  \nonumber \\ &-&
     2\,g^2_R\,M_R\,\cos (\Phi_1 + \Phi_2 - \Phi_3)\,
      |v_{\Delta_R }|\,|v_{\delta_R }|\,|\mu_3 |+
     2\,{|{\tilde M}_R |}^2\,\left( g^2_R\,(v_{\Delta_R})^2 +
        {|\mu_3 |}^2 \right) \,{\sin (\theta_\kappa)}^2 \nonumber \\
        &-&
     m^2_{{\tilde \chi}^\pm_j}\,\left( M^2_R + g^2_R\,
         \left( (v_{\Delta_R})^2 + (v_{\delta_R})^2 \right)  +
        {|\mu_3 |}^2 + 2\,{|{\tilde M}_R |}^2\,{\sin (\theta_\kappa)}^2
        \right)  \biggr]  +  \sin (\theta_\kappa) \nonumber \\ &\times&
   \biggl[ \mu \,\biggl(  2\,g^2_R\,M_R\,
         \cos (\Phi_1 + \Phi_2 - \Phi_3)\,|v_{\Delta_R }|\,
         |v_{\delta_R }|\,|\mu_3 | - {m_{{\tilde \chi}^\pm_j}}^4 - g^4_R\,(v_{\Delta_R})^2\,
         (v_{\delta_R})^2  \nonumber \\&-&
        M^2_R\,{|\mu_3 |}^2 +
        m^2_{{\tilde \chi}^\pm_j}\,\left( M^2_R + g^2_R\,
            \left( (v_{\Delta_R})^2 + (v_{\delta_R})^2 \right)  +
           {|\mu_3 |}^2 \right)  \biggr)  +
     e^{-i \left(\Phi_1 + \Phi_2 \right) }\,{\tilde M}^2_R\, \nonumber \\ &\times&  \,\sin (2\,\theta_\kappa)
      \biggl( e^{i \,\Phi_3}\,g^2_R\,|v_{\Delta_R }|\,
         |v_{\delta_R }|\,|\mu_3 | +
        e^{i \,\left( \Phi_1 + \Phi_2 \right) }\,M_R\,
         \left( m^2_{{\tilde \chi}^\pm_j} - {|\mu_3 |}^2 \right)  \biggr)
         \biggr]
\biggr\} \label{eq:pj4} \eeqa \beqa p_{j5}&=& {- \, 2 \, g_R {\tilde
M}^\ast_L \, {\tilde M}^\ast_R \, e^{-i(\Phi_2+ \Phi_\mu)} \, \over
{\tilde {\cal D}}_j} \biggl\{ e^{ 2\,i \,{\tilde
\Phi}_L}\,{\sqrt{\frac{P^{U^\ast}_{j11}}{P^{V}_{j11}}}}\,\xi^\ast_j\,
   \cos (\theta_\kappa)\,m_{{\tilde \chi}^\pm_j}\,\biggl[ e^{ 2\,i \,{\tilde \Phi}_R}\,|\mu |\,\sin (\theta_\kappa)
       \nonumber \\ &\times& \left( e^{i \,\left( \Phi_1 + \Phi_2 \right) }\,M_R\,
           v_{\Delta_R} + e^{i \,\Phi_3}\,v_{\delta_R}\,
           |\mu_3 | \right) -
     e^{i \,\Phi_\mu}\,\cos (\theta_\kappa)\,
      \biggl( e^{i \,\Phi_3}\,M_R\,v_{\delta_R}\,
         |\mu_3 | \nonumber \\ &+& e^{i \,\left( \Phi_1 + \Phi_2 \right) }\,
         v_{\Delta_R}\,  \left( m^2_{{\tilde \chi}^\pm_j} - g^2_R\,(v_{\delta_R})^2 -
           2\,{|{\tilde M}_R |}^2\,{\sin (\theta_\kappa)}^2 \right)  \biggr)
           \biggr] +  e^{i \,\Phi_\mu}\,\sin (\theta_\kappa)\,
           \nonumber \\ &\times&
   \biggl[ e^{i \,\Phi_\mu}\,\cos (\theta_\kappa)\,|\mu |\,
      \biggl( e^{i \,\left( \Phi_1 + \Phi_2 \right) }\,v_{\Delta_R}\, \left( m^2_{{\tilde \chi}^\pm_j} - g^2_R\,(v_{\delta_R})^2
\right)  +  e^{i \,\Phi_3}\,M_R\,v_{\delta_R}\,|\mu_3 |
        \biggr) \nonumber \\ &-& e^{2\,i  \,{\tilde \Phi}_R}\, \sin (\theta_\kappa)
      \biggl( e^{i \,\left( \Phi_1 + \Phi_2 \right) }\,M_R\,m^2_{{\tilde \chi}^\pm_j}\,
           v_{\Delta_R}  +
        e^{i \,\Phi_3}\, v_{\delta_R}\,
         |\mu_3 | \, \nonumber \\ &\times& \left( m^2_{{\tilde \chi}^\pm_j} -
           2\,{\cos (\theta_\kappa)}^2\,{|{\tilde M}_R |}^2 \right) \biggr)  \biggr] \biggr\} \label{eq:pj5}\eeqa

 The reduced projectors ${\tilde p}_{j2}, {\tilde p}_{j3}$ and ${\tilde p}_{j5}$
                  are obtained by interchanging $v_{\Delta_R} \leftrightarrow v_{\delta_R},$
$\Phi_1 \leftrightarrow  \Phi_2$ and $\sin \theta_\kappa
\leftrightarrow \cos\theta_\kappa$ in
                  $p^\ast_{j2}, p^\ast_{j4}$ and $ p^\ast_{j5},$
                  respectively,  On the other hand, taking into account the
analysis at the end of Section \ref{sec-diagonalizing}, we deduce
that $p_{\alpha 3}= p_{3 \alpha}=0, \alpha=1,2,4,5.$ Similarly,
${\tilde p}_{\alpha 4}=0, \, \alpha= 1,2,4,5,$ and $ {\tilde p}_{3
\beta}, \, \beta=1,2,3,5.$ Moreover, without any loss of generality
we can chose $p_{33}={\tilde p}_{34}=1.$

\subsection{Connection with the Jarlskog's formulation}

Combining  Eq. \eqref{eq:CPjalpha-V} with Eq. \eqref{eq:sysV} and
Eq. \eqref{eq:CPjalpha-U} with  \eqref{eq:tsysU},  it is easy to
verify that the $V$-type and $U^\ast$-type projectors satisfies  the
standard projector relations \be P_i  \, P_j = P_j \, \delta_{ij},
\qquad {\rm Tr} P_j =1, \qquad \sum_{j=1}^5 P_j = I_5, \qquad
P_{j\alpha \alpha} P_{j \beta\beta} = |P_{j \alpha \beta}|^2. \ee

 In general,
in  the Jarlskog's formulation \cite{Jarlskog}, the  $P^V_j$
projectors writes ($j=1,\ldots,5$) \be P^V_j = \prod_{k=1; \, k \ne
j}^{5} {m^2_{{\tilde \chi}^\pm_k} - H \over  m^2_{{\tilde
\chi}^\pm_k} - m^2_{{\tilde \chi}^\pm_j}}. \ee We can show that they
also can  be written in the form \be \label{eq:PV-DJ}  P^V_j =
{{\tilde P}^V_j \over {\tilde \Delta}_j}, \ee where \be {\tilde
\Delta}_j = 4 (m^2_{{\tilde \chi}^\pm_j})^5 - 3 {\tilde a} \,
(m^2_{{\tilde \chi}^\pm_j})^4 + 2 {\tilde b} \, (m^2_{{\tilde
\chi}^\pm_j})^3 - {\tilde c} \, (m^2_{{\tilde \chi}^\pm_j})^2 +
{\tilde e} \ee and

\beqa \nonumber {\tilde P}^V_{j \alpha \beta} &=& (m^2_{{\tilde
\chi}^\pm_j})^4 \,  H_{\alpha \beta} + (m^2_{{\tilde \chi}^\pm_j})^3
(H^2_{\alpha \beta} - {\tilde a} \,  H_{\alpha \beta}) +
(m^2_{{\tilde \chi}^\pm_j})^2 ( H^3_{\alpha \beta} - {\tilde a} \,
H^2_{\alpha \beta} + {\tilde b} \,  H_{\alpha \beta}) \nonumber \\
&+& m^2_{{\tilde \chi}^\pm_j} ( H^4_{\alpha \beta} - {\tilde a} \,
H^3_{\alpha \beta} + {\tilde b} \,  H^2_{\alpha \beta} - {\tilde c}
\, H_{\alpha \beta}) + {\tilde e} \, \delta_{\alpha \beta}.
\label{eq:tildePVjalphabeta}\eeqa Here,  ${\tilde a}, \ldots,
{\tilde e}$ are the coefficients of the characteristic polynomial
which determine  the chargino masses\be \label{eq:poli-caract-total}
 {\rm Det} \left(m^2_{{\tilde \chi}^\pm_j} \, I_5  - H \right) = (m^2_{{\tilde \chi}^\pm_j})^5 - {\tilde a} \, (m^2_{{\tilde \chi}^\pm_j})^4 +  {\tilde b} \,  (m^2_{{\tilde \chi}^\pm_j})^3
 - {\tilde c} \,  (m^2_{{\tilde \chi}^\pm_j})^2 + {\tilde d} \,  (m^2_{{\tilde \chi}^\pm_j}) - {\tilde e} , \ee i.e., comparing Eq. \eqref{eq:CEQ}  with Eq.
 \eqref{eq:poli-caract-total}, the coefficients  are given by
 \be
{\tilde a} = a + |\mu|^2, \quad  {\tilde b} = b + a \, |\mu|^2,
\quad {\tilde c} = c + b \,  |\mu|^2,  \quad  {\tilde d} = d + c \,
|\mu|^2, \quad  {\tilde e} = d  \, |\mu|^2.
 \ee
Now, inserting \eqref{eq:PV-DJ} into \eqref{eq:CPjalpha-V} and using
the definition \eqref{eq:def-red-proj-v} for the type-$V$ reduced
projectors, we can show that ($\alpha=1,\ldots,5$) \be
\label{eq:Gen-ptilde-delta-V} {\tilde P}^{V}_{j \ell \alpha} =
m_{{\tilde \chi}^{\pm}_{j}}^2 \, \,
 {(\Delta^{(\ell)}_{\alpha j})}^\ast.\ee

Similarly, for the $U^\ast$-type projectors we have
($\alpha=1,\ldots,5$ )
 \be \label{eq:Gen-ptilde-delta-U} {\tilde P}^{U^\ast}_{j \ell
\alpha} = m_{{\tilde \chi}^{\pm}_{j}}^2 \, \,
 {({\tilde \Delta}^{(\ell)}_{\alpha j})}^\ast ,  \ee
where the quantities ${\tilde P}^{U^\ast}_{j \alpha \beta}$ are
given by \eqref{eq:tildePVjalphabeta} but with ${\tilde H}$ in the
place of $H.$

Both, \eqref{eq:Gen-ptilde-delta-V} and  Eq.
\eqref{eq:Gen-ptilde-delta-U} satisfy identically when
$\alpha=1,\ldots,5, \, \alpha \ne \ell,$ whereas when $\alpha=\ell,$
they are useful equivalences.

\subsection{Explicit form of some relevant projectors}
As an example, let us compute the projectors ${\tilde
P}^V_{j\ell\ell}$ and ${\tilde P}^{U^\ast}_{j\ell\ell},$ when
$\ell=1.$ From equation \eqref{eq:tildePVjalphabeta} with
$\alpha=\beta=1,$ we can show that the  $V$-type quantities ${\tilde
P}^V_{j11}$ can be written in the form \be {\tilde P}^V_{j11} =
\left( m^2_{{\tilde \chi}^\pm_j} - |\mu|^2 \right) \, \left[ {\tilde
{\cal D}}_j \, |M_L |^2 + {\tilde {\cal B}}_j |M_L| + {\tilde {\cal
C}}^V_j \right] , \label{eq:second-order-tpv}\ee where \beqa {\tilde
{\cal D}}_j &=& m^6_{{\tilde \chi}^\pm_j} -
     g^4_R\,(v_{\Delta_R})^2\,(v_{\delta_R})^2\,
      {|\mu |}^2 - m^4_{{\tilde \chi}^\pm_j}\,\left( M^2_R + 2\,{|{\tilde M}_R |}^2 +
        g^2_R\,\left( (v_{\Delta_R})^2 + (v_{\delta_R})^2 \right)  +
        {|\mu |}^2 + {|\mu_3 |}^2 \right)  \nonumber \\ &+&
     2\,g^2_R\,v_{\Delta_R}\,v_{\delta_R}\,|\mu |\,
      |\mu_3 |\,\left( M_R\,
         \cos ( \Theta_1 ) \,|\mu | +
        \cos (\Theta_1 - \Theta_2)\,
         {|{\tilde M}_R |}^2\,\sin (2\,\theta_\kappa) \right) \nonumber \\ &-&
     {|\mu_3 |}^2\,\left( M^2_R\,{|\mu |}^2 +
        2\,M_R\,\cos ( \Theta_2 ) \,{|{\tilde M}_R |}^2\,
         |\mu |\,\sin (2\,\theta_\kappa) +
        {|{\tilde M}_R |}^4\,{\sin (2\,\theta_\kappa)}^2 \right)  \nonumber \\ &+&
     m^2_{{\tilde \chi}^\pm_j}\,\biggr( g^4_R\,(v_{\Delta_R})^2\,(v_{\delta_R})^2 +
        M^2_R\,{|\mu |}^2 +
        g^2_R\,\left( (v_{\Delta_R})^2 + (v_{\delta_R})^2 \right) \,
         {|\mu |}^2 - 2\,g^2_R\,M_R\,
         \cos ( \Theta_1 ) \,v_{\Delta_R}\,
         v_{\delta_R}\,|\mu_3 | \nonumber \\ &+&
        \left( M^2_R + {|\mu |}^2 \right) \,{|\mu_3 |}^2 +
        {|{\tilde M}_R |}^4\,{\sin (2\,\theta_\kappa)}^2 \nonumber \\ &+&
        2\,{|{\tilde M}_R |}^2\,\left( {|\mu_3 |}^2 +
           g^2_R\,\left( {\cos (\theta_\kappa)}^2\,(v_{\delta_R})^2 +
              (v_{\Delta_R})^2\, \sin^2 (\theta_\kappa) \right)  +
           M_R\,\cos ( \Theta_2 ) \,|\mu |\,
            \sin (2\,\theta_\kappa) \right)  \biggl), \label{eq:calDtildej}\eeqa

\beqa {\tilde {\cal B}}_j &=& 2 \,  {|{\tilde M}_L |}^2\,\sin
(2\,\theta_\kappa)\,
   \Biggl[ \cos (\Theta_3 - \Theta_2)\,|\mu |\,
  \nonumber \\ &\times&
      \biggl( -m^4_{{\tilde \chi}^\pm_j} - g^4_R\,(v_{\Delta_R})^2\,(v_{\delta_R})^2 +
        2\,g^2_R\,M_R\,\cos ( \Theta_1 ) \,
         v_{\Delta_R}\,v_{\delta_R}\,|\mu_3 | -
        M^2_R\,{|\mu_3 |}^2 \nonumber \\ &+&
        m^2_{{\tilde \chi}^\pm_j}\,\left( M^2_R + g^2_R\,
            \left( (v_{\Delta_R})^2 + (v_{\delta_R})^2 \right)  +
           {|\mu_3 |}^2 \right)  \biggr)  \nonumber \\ &+&
     {|{\tilde M}_R |}^2\,\biggl( g^2_R\,
         \cos (\Theta_1 - \Theta_3) \,v_{\Delta_R}\,v_{\delta_R}\,
         |\mu_3 | \nonumber \\ &+& M_R\,
         \cos ( \Theta_3 ) \,
         \left( m^2_{{\tilde \chi}^\pm_j} - {|\mu_3 |}^2 \right)  \biggr) \,\sin (2\,\theta_\kappa)
         \Biggr],\eeqa

\beqa {\tilde {\cal C}}^V_j &=& {|{\tilde M}_L |}^2\,\Biggl[
-2\,m^2_{{\tilde \chi}^\pm_j}\,
      \biggl( -m^4_{{\tilde \chi}^\pm_j} - g^4_R\,(v_{\Delta_R})^2\,(v_{\delta_R})^2 +
        2\,g^2_R\,M_R\,\cos ( \Theta_1 ) \,
         v_{\Delta_R}\,v_{\delta_R}\,|\mu_3 | \nonumber \\ &-&
        M^2_R\,{|\mu_3 |}^2 -
        2\,{\cos (\theta_\kappa)}^2\,{|{\tilde M}_R |}^2\,
         \left( g^2_R\,(v_{\delta_R})^2 + {|\mu_3 |}^2 \right)  \nonumber \\ &+&
        m^2_{{\tilde \chi}^\pm_j}\,\left( M^2_R + 2\,{\cos (\theta_\kappa)}^2\,
            \left( {|{\tilde M}_L |}^2 + {|{\tilde M}_R |}^2 \right)  +
           g^2_R\,\left( (v_{\Delta_R})^2 + (v_{\delta_R})^2 \right)  +
           {|\mu_3 |}^2 \right)  \biggr) \,{\sin (\theta_\kappa)}^2  \nonumber \\
           &+&
     {|{\tilde M}_L |}^2\,\bigl( -\left( g^4_R\,(v_{\Delta_R})^2\,
           (v_{\delta_R})^2 \right)  +
        2\,g^2_R\,M_R\,\cos ( \Theta_1 ) \,
         v_{\Delta_R}\,v_{\delta_R}\,|\mu_3 | -
        M^2_R\,{|\mu_3 |}^2 \nonumber \\ &+&
        m^2_{{\tilde \chi}^\pm_j}\,\left( M^2_R + g^2_R\,
            \left( (v_{\Delta_R})^2 + (v_{\delta_R})^2 \right)  +
           {|\mu_3 |}^2 \right)  \bigr) \,{\sin (2\,\theta_\kappa)}^2
           \Biggr].\eeqa

Similarly, the $U^\ast$-type quantities  ${\tilde P}^{U^\ast}_{j11}$
can be written in the form \be {\tilde P}^{U^\ast}_{j11} = \left(
m^2_{{\tilde \chi}^\pm_j} - |\mu|^2 \right) \, \left[ {\tilde {\cal
D}}_j \, |M_L |^2 + {\tilde {\cal B}}_j |M_L| + {\tilde {\cal
C}}^{U^\ast}_j \right] , \ee where the coefficients $ {\tilde {\cal
C}}^{U^\ast}_j $ are obtained by interchanging  $v_{\Delta_R}
\leftrightarrow v_{\delta_R}$  and $\sin \theta_\kappa
\leftrightarrow \cos
                 \theta_\kappa$ in $ {\tilde {\cal C}}^{V}_j.$

\section{Parameter problem determination}
\label{sec-det-parameters} In this section we implement a procedure
to express the parameter $M_L$ and the phase angle $\Theta_3$ in
terms of  chargino masses,  eigenphases, and complementary set of
fundamental parameters. In fact, each of the fundamental parameters
which are located on the diagonal of the chargino mass matrix $M$
can be easily disentangled from the rest the parameters using the
generalized projector formalism.  In the $M$ matrix diagonal we have
the complex parameter $M_L,$ the real parameter $M_R$ and the
complex parameter $\mu_3.$ To disentangle $M_L$ we have to use the
generalized projector formulation with $\ell=1,$ whereas to
disentangle $M_R$ and $\mu_3$ we have to take $\ell=2$ and $\ell=5,$
respectively.

\subsection{\boldmath $M_L$ in terms of $m_{{\tilde \chi}^\pm_j},$ $\zeta_j,$  and the complementary parameters}

To express the fundamental parameter $M_L$ in terms of the physical
chargino masses, the eigenphases, and the remaining parameters we
can use either Eq. \eqref{eq:gen-inversionU} or Eq.
\eqref{eq:gen-inversionV}, with $\alpha=\ell=1.$ Thus, for instance,
from Eq. \eqref{eq:gen-inversionU} for $\alpha=1,$ considering that
${\tilde p}_{j1}=1, j=1,2,4,5$  and using \eqref{eq:Mmatrix}, we get
\be   m_{{\tilde \chi}^{\pm}_{j}} \, \zeta_j \, \sqrt{{\tilde
P}^{U^\ast}_{j11} \over {\tilde P}^V_{j11}} = M_{L} + \sqrt{2}
{\tilde M}_L \cos\theta_k p^\ast_{j4} . \label{eq:xij-mj-pj} \ee Now
we have two ways to continue. We can either substitute Eq.
\eqref{eq:pj4} into \eqref{eq:xij-mj-pj} or express  $p^\ast_{j4}$
in terms of $\Delta_{1j}$ and $\Delta_{4j},$ according to  Eq.
\eqref{eq:def-red-proj-v}. Then using the equivalences
\eqref{eq:Gen-ptilde-delta-V} and \eqref{eq:Gen-ptilde-delta-U}, we
get \be  M_{L} = {\tilde A}_j \, \zeta_j + {\tilde B}_j, \qquad
j=1,2,4,5 \label{genMLformula}\ee where \be {\tilde A}_j= -
{\sqrt{{\tilde \Delta}_{1j} \, \Delta_{1j}} \over (m_{{\tilde
\chi}^{\pm}_{j}}^2 - |\mu|^2)} {m_{{\tilde \chi}^{\pm}_{j}} \over
{\tilde {\cal D}}_j}, \label{eq:tildeAJ}\ee and \beqa \nonumber
{\tilde B}_j&=& { {\tilde M}_L^2 \sin(2\theta_k) \over {\tilde {\cal
D}}_j}\biggl[ \mu \,\biggl(  2\,g^2_R\,M_R\,
         \cos (\Phi_1 + \Phi_2 - \Phi_3)\,|v_{\Delta_R }|\,
         |v_{\delta_R }|\,|\mu_3 | - {m_{{\tilde \chi}^\pm_j}}^4 - g^4_R\,(v_{\Delta_R})^2\,
         (v_{\delta_R})^2  \nonumber \\&-&
        M^2_R\,{|\mu_3 |}^2 +
        m^2_{{\tilde \chi}^\pm_j}\,\left( M^2_R + g^2_R\,
            \left( (v_{\Delta_R})^2 + (v_{\delta_R})^2 \right)  +
           {|\mu_3 |}^2 \right)  \biggr)  +
     e^{-i \left(\Phi_1 + \Phi_2 \right) }\,{\tilde M}^2_R\, \nonumber \\ &\times&  \,\sin (2\,\theta_\kappa)
      \biggl( e^{i \,\Phi_3}\,g^2_R\,|v_{\Delta_R }|\,
         |v_{\delta_R }|\,|\mu_3 | +
        e^{i \,\left( \Phi_1 + \Phi_2 \right) }\,M_R\,
         \left( m^2_{{\tilde \chi}^\pm_j} - {|\mu_3 |}^2 \right)  \biggr)
         \biggr],\label{eq:tildeBJ}\eeqa where ${\tilde {\cal D}}_j$ is given by
Eq. \eqref{eq:calDtildej}.

Equation \eqref{genMLformula} allows us to determinate the behaviour
of $|M_L|$ and $\Phi_L$ in terms of the eigenphases $\zeta_j$ and
the physical masses $m_{{\tilde \chi}^{\pm}_{j}},$ when the rest of
fundamental parameters are known. This is a general equation which
take into account the contribution of the right handed Higgisino
fields.

\subsection{Disentangling \boldmath $|M_L|$} It is  useful to express the norm of $M_L$ in terms of the physical masses and remaining parameters, without any explicit dependence
on the eigenphases.  We combine Eq. \eqref{eq:Gen-ptilde-delta-V},
when $\alpha=\ell=1,$ with  Eqs. \eqref{eq:delta1j} and
\eqref{eq:second-order-tpv} to get the quadratic equation
 \be {\tilde {\cal D}}_j \, |M_L|^2 + {\tilde {\cal
B}}_j \, |M_L|+ {\tilde {\cal C}}_j =0, \qquad j=1,2,4,5, \ee where
${\tilde {\cal C}}_j = {\tilde {\cal C}}^V_j - \Delta_{1j} /
(m^2_{{\tilde \chi}^\pm_j} - |\mu|^2).$ Solving for $|M_L|$ we get
\be |M_L|= {- {\tilde {\cal B}}_j \pm \sqrt{ {\tilde {\cal B}}^2_j -
4 \, {\tilde {\cal C}}_j \, {\tilde {\cal D}}_j} \over 2 \, {\tilde
{\cal D}}_j }. \label{eq:MLNORM} \ee  Equation \eqref{eq:MLNORM}
expresses the norm of $M_L$ in terms of the physical chargino
masses, the CP-violating phases and the remaining parameters. Since
 $|M_L| \ge 0,$ the following constraints must be satisfied: \be
{\tilde {\cal B}}^2_j - 4 \, {\tilde {\cal C}}_j \, {\tilde {\cal
D}}_j \ge 0, \qquad {{\tilde {\cal B}}_j \over {\tilde {\cal D}}_j}
< 0.\ee

These results include the contribution of the right handed Higgino
parameters.

\subsection{\boldmath $|M_L|$ in terms of the two lightest chargino masses}

Writing  ${\tilde {\cal B}}_j $  in the form \be
\label{eq:tildecalBj}
    {\tilde {\cal B}}_j = {{\tilde {\cal P}}_j + {\tilde {\cal Q}}_j
    \tan (\Theta_3) \over \sqrt{1 + \tan^2 (\Theta_3) }}, \qquad
    j=1,2,
\ee with \beqa {\tilde {\cal P}}_j &=& 2 \,  {|{\tilde M}_L
|}^2\,\sin (2\,\theta_\kappa)\, \Biggl[ |\mu |\, \cos (\Theta_2) \,
\biggl( m^2_{{\tilde \chi}^\pm_j}\,\left( M^2_R + g^2_R\,
            ( (v_{\Delta_R})^2 + (v_{\delta_R})^2)  +
           {|\mu_3 |}^2 \right)
  \nonumber \\ &-&
       m^4_{{\tilde \chi}^\pm_j} - g^4_R\,(v_{\Delta_R})^2\,(v_{\delta_R})^2 +
        2\,g^2_R\,M_R\,\cos ( \Theta_1 ) \,
         v_{\Delta_R}\,v_{\delta_R}\,|\mu_3 | -
        M^2_R\,{|\mu_3 |}^2 \biggr)
        \nonumber \\ &+&
     {|{\tilde M}_R |}^2\, \sin (2\,\theta_\kappa) \biggl( g^2_R\,
         \cos (\Theta_1) \,v_{\Delta_R}\,v_{\delta_R}\,
         |\mu_3 | +  M_R  \,
         ( m^2_{{\tilde \chi}^\pm_j} - {|\mu_3 |}^2 )  \biggr)
         \Biggr] \eeqa and  \beqa {\tilde {\cal Q}}_j &=& 2 \,  {|{\tilde M}_L |}^2\,\sin
(2\,\theta_\kappa)\, \Biggl[ |\mu |\, \sin (\Theta_2) \, \biggl(
m^2_{{\tilde \chi}^\pm_j}\,\left( M^2_R + g^2_R\,
            ( (v_{\Delta_R})^2 + (v_{\delta_R})^2)  +
           {|\mu_3 |}^2 \right)
  \nonumber \\ &-&
       m^4_{{\tilde \chi}^\pm_j} - g^4_R\,(v_{\Delta_R})^2\,(v_{\delta_R})^2 +
        2\,g^2_R\,M_R\,\cos ( \Theta_1 ) \,
         v_{\Delta_R}\,v_{\delta_R}\,|\mu_3 | -
        M^2_R\,{|\mu_3 |}^2 \biggr)
        \nonumber \\ &+&
     {|{\tilde M}_R |}^2\, \sin (2\,\theta_\kappa) \,  g^2_R\,
         \sin (\Theta_1) \,v_{\Delta_R}\,v_{\delta_R}\,
         |\mu_3 |
         \Biggr],\eeqa
and inserting it into Eq. \eqref{eq:MLNORM}, after some algebraic
manipulations we get \be \label{eq:THETA3} \tan (\Theta_3) = {\tilde
{\mathbb R}}\equiv  { - {\tilde {\mathbb B}} +  {\tilde \epsilon}
\sqrt{{\tilde {\mathbb B}}^2 - 4 {\tilde {\mathbb A}} {\tilde
{\mathbb C}}} \over 2 {\tilde {\mathbb A}} },\ee where ${\tilde
{\mathbb B}}^2 - 4 {\tilde {\mathbb A}} {\tilde {\mathbb C}} \ge 0,$
\be \label{eq:tildematA} {\tilde {\mathbb A}} = {1 \over 2}
F({\tilde {\cal Q}}_1,{\tilde {\cal Q}}_2,{\tilde {\cal
Q}}_1,{\tilde {\cal Q}}_2) - ({\tilde {\cal D}}_1 {\tilde {\cal
C}}_2 - {\tilde {\cal D}}_2 {\tilde {\cal C}}_1)^2,\ee \be
\label{eq:tildematB} {\tilde {\mathbb B}} = F({\tilde {\cal
P}}_1,{\tilde {\cal P}}_2,{\tilde {\cal Q}}_1,{\tilde {\cal Q}}_2)
\ee and \be \label{eq:tildematC} {\tilde {\mathbb C}} = {1 \over 2}
F({\tilde {\cal P}}_1,{\tilde {\cal P}}_2,{\tilde {\cal
P}}_1,{\tilde {\cal P}}_2) - ({\tilde {\cal D}}_1 {\tilde {\cal
C}}_2 - {\tilde {\cal D}}_2 {\tilde {\cal C}}_1)^2,\ee with \beqa
\nonumber  F({\tilde {\cal P}}_1,{\tilde {\cal P}}_2,{\tilde {\cal
Q}}_1,{\tilde {\cal Q}}_2) &=&  ({\tilde {\cal D}}_1 {\tilde {\cal
C}}_2 + {\tilde {\cal D}}_2 {\tilde {\cal C}}_1) ({\tilde {\cal
P}}_1 {\tilde {\cal Q}}_2 + {\tilde {\cal P}}_2 {\tilde {\cal Q}}_1)
\\\nonumber &-& 2 ( {\tilde {\cal D}}_1 {\tilde {\cal C}}_1 {\tilde
{\cal P}}_2 {\tilde {\cal Q}}_2  + {\tilde {\cal D}}_2
{\tilde {\cal C}}_2 {\tilde {\cal P}}_1 {\tilde {\cal Q}}_1),\label{eq:FP1P2Q1Q2} \\
\eeqa where $\tilde \epsilon=\pm 1.$

Moreover, combining  Eq. \eqref{eq:MLNORM} for $j=1$ with Eq.
\eqref{eq:MLNORM} for  $j=2,$ and then using Eq.
\eqref{eq:tildecalBj} for $j=1,2,$ we get \be
\label{eq:ML2charginos}|M_L|={ ({\tilde {\cal D}}_1 {\tilde {\cal
C}}_2 - {\tilde {\cal D}}_2 {\tilde {\cal C}}_1)\, \sqrt{1 + {\tilde
{\mathbb R}}^2} \over ({\tilde {\cal D}}_2 {\tilde {\cal P}}_1 -
{\tilde {\cal D}}_1 {\tilde {\cal P}}_2 )+ ({\tilde {\cal D}}_2
{\tilde {\cal Q}}_1 - {\tilde {\cal D}}_1 {\tilde {\cal Q}}_2)
{\tilde {\mathbb R}}}.\ee Equations \eqref{eq:THETA3} and
\eqref{eq:ML2charginos} allow us to determine the phase $\Theta_3 $
and the norm $|M_L|,$ respectively, up to a twofold discrete
ambiguity, in terms of the two lightest chargino masses and the
remaining fundamental parameters.

\section{Complete parameter inversion}
\label{sec-par-inversion} The  fundamental parameter reconstruction
from measurements of some suitable physical observables is a non
trivial problem in many SUSY theories beyond the Standard Model.  In
the context of the MSSM, some important techniques have been
introduced in the literature to obtain  the chargino and neutralino
parameters. For instance, measurements of some cross section type
observables, involving the neutralino and chargino pair production
in electron-positron annihilation
 \cite{MSSM-det-par-1,MSSM-det-par-2,MSSM-det-par-3,MSSM-det-par-4,MSSM-det-par-5,MSSM-det-par-6,MSSM-det-par-7,MSSM-det-par-8}.
Using this principle and the projector technique, it is possible to
implement a systematic method to reconstruct the fundamental
neutralino and chargino parameters from the experiments, in the
context of either the MSSM or the LRSUSY model
\cite{key12,key13,key11,nalvar8}.

Let us consider some class of cross section-type observales
associated with  the chargino pair production ${\tilde \chi_i}^\pm
{\tilde \chi_j}^\pm, \, i,j=1,2,\ldots,5, $ from the $e^+ e^-$
annihilation, at the future Linear Collider. For instance, total
cross section with polarized or unpolarized beams, angular chargino
distribution, forward and backward asymmetries. These type of
observables depend, in addition to the chargino masses and leptons
masses, on the entries of the $V$ and $U^\ast$
\cite{Frank-neu-char-masses-2,cross-section-LRSUSY}. Thus, in
principle, if we are able to measure a set of appropriate cross
section-type observables and to determine the entries of $V$ and
$U,$ then we could invert Eqs. \eqref{eq:GEN-EVEVij} and
\eqref{eq:GEN-EVEUij} to find the fundamental LRSUSY parameters.
However, at this stage, the inversion process is very difficult
because the high complexity of these relations. It is necessary to
find out a new parametrization of the cross section-type
observables. A more suitable procedure is parameterizing these
observables in terms of the fundamental reduced projectors and
eigenphases. It can be done easily by using Eqs. \eqref{eq:Vaj} and
\eqref{eq:Uaj}. Then, if we are able to determine the value of the
reduced projectors and eigenphases from the experiments, with the
help of the Eqs. \eqref{eq:gen-inversionU} and
\eqref{eq:gen-inversionV}, we can find appropriate expressions to
reconstruct the fundamental parameters.

\subsection{Fundamental parameter inversion equations}
Using the basic Eqs. \eqref{eq:gen-inversionU} and
\eqref{eq:gen-inversionV}, we can express the fundamental parameters
in terms of the reduced projectors, the physical masses and the
eigenphases.  We get (j=1,2,4,5)
 \be \label{eq:ML-pj-etaj-thetaj} M_L = m_{{\tilde
\chi}^{\pm}_{j}} \, \zeta_j \, \, { \sqrt{P^{U^\ast}_{j11}\over
P^V_{j11}} {\tilde p}_{j3} \tan\theta_k - \sqrt{P^V_{j11} \over
P^{U^\ast}_{j11}} p_{j4}^\ast
  \over {\tilde p}_{j3} \tan\theta_k - p_{j4}^\ast},\ee \be \label{eq:MtL-pj-etaj-thetaj} {\tilde M}_L = - { m_{{\tilde \chi}^{\pm}_{j}} \,
\zeta_j \over \sqrt{2}} \, \, {\sqrt{P^{U^\ast}_{j11}\over
P^V_{j11}} - \sqrt{P^V_{j11} \over P^{U^\ast}_{j11}}  \over {\tilde
p}_{j3} \sin\theta_k - p_{j4}^\ast \cos\theta_k },   \ee \be
\label{eq:MtR-pj-etaj-thetaj} {\tilde M}_R = { m_{{\tilde
\chi}^{\pm}_{j}} \,   \zeta_j \over \sqrt{2}} \, \,
{\sqrt{P^{U^\ast}_{j11}\over P^V_{j11}} (|{\tilde p}_{j2}|^2 +
|{\tilde p}_{j5}|^2 ) - \sqrt{P^V_{j11} \over P^{U^\ast}_{j11}} (
|p_{j2}|^2 + |p_{j5}|^2) \over p^\ast_{j4} {\tilde p}_{j2}
\cos\theta_k - p^\ast_{j2} {\tilde p}_{j3} \sin\theta_k }, \ee \beqa
\mu &=& {m_{{\tilde \chi}^{\pm}_{j}} \, \zeta_j \over ( p_{j4}^\ast
- {\tilde p}_{j3} \tan\theta_k ) ( {\tilde p}_{j3} p^\ast_{j2}
\tan\theta_k -  {\tilde p}_{j2} p_{j4}^\ast)} \nonumber  \\
&\times& \biggl\{ \sqrt{P^{U^\ast}_{j11}\over P^V_{j11}} \left[
{\tilde p}_{j2} {\tilde p}_{j3}^\ast p_{j4}^\ast  - \left(
(p^\ast_{j2} - {\tilde p}_{j2}) (|{\tilde p}_{j2}|^2 + |{\tilde
p}_{j5}|^2)+ p^\ast_{j2} |{\tilde p}_{j3}|^2\right) \tan\theta_k
\right] \nonumber \\ &+& \sqrt{P^{V}_{j11}\over P^{U^\ast}_{j11}}
\tan\theta_k \left[ p_{j2}^\ast {\tilde p}_{j3} p_{j4} \tan\theta_k
- \left( ({\tilde p}_{j2} - p^\ast_{j2}) (|p_{j2}|^2 + |p_{j5}|^2) +
{\tilde p}_{j2} |p_{j4}|^2\right)\right] \biggr\},
\label{eq:mu-pj-etaj-thetaj} \eeqa \be \sqrt{2} M_1
\sin\theta\upsilon
 = g_R \,  v_{\Delta_R} e^{i \Phi_1} = { m_{{\tilde
\chi}^\pm_j } \zeta_j \, \sqrt{P^{V}_{j11}\over P^{U^\ast}_{j11}} \,
p_{j5}  - \mu_3 \, {\tilde p}_{j5}
  \over {\tilde p}_{j2}}, \label{eq:v-Delta-R-M1}
 \ee  \be \sqrt{2} M_2 \cos\theta\upsilon = g_R \, v_{\delta_R} e^{i \Phi_2} = { m_{{\tilde \chi}^\pm_j } \zeta_j \, \sqrt{P^{U^\ast}_{j11}\over P^{V}_{j11}}  \, {\tilde p}^\ast_{j5}  - \mu_3 \, p^\ast_{j5}
  \over  p^\ast_{j2}} \label{eq:v-delta-R-M2}
 \ee and
\beqa \nonumber  M_R &=&\mu_3 \, { p^\ast_{j5} \, {\tilde p}_{j5}
\over p^\ast_{j2} {\tilde p}_{j2}} + { m_{{\tilde \chi}^\pm_j }
\zeta_j  \over p_{j2}^\ast {\tilde p}_{j3} \tan\theta_k - {\tilde
p}_{j2} p_{j4}^\ast} \biggl[ \sqrt{P^{U^\ast}_{j11}\over P^V_{j11}}
{\tilde p}_{j2}^\ast {\tilde p}_{j3} \tan\theta_k - \sqrt{P^V_{j11}
\over P^{U^\ast}_{j11}} p_{j2} p_{j4}^\ast
   \\ &+&  { \sqrt{P^{U^\ast}_{j11} \over P^{V}_{j11}}  p^\ast_{j4} \, {\tilde p}_{j2}
   |{\tilde p}_{j5}|^2 -  \sqrt{P^{V}_{j11} \over
P^{U^\ast}_{j11}} {\tilde p}_{j3} \, p^\ast_{j2} \,  |p_{j5}|^2
\tan\theta_k \over p^\ast_{j2} {\tilde p}_{j2}} \biggr].
 \label{eq:MR-pj-etaj-thetaj} \eeqa

When $j=3,$   we only   have \be \zeta_3 =  e^{i \vartheta_3}= -
{\mu \over |\mu|}. \label{eq:zetaj3} \ee

Note that in  Eqs. \eqref{eq:v-Delta-R-M1} and
\eqref{eq:v-delta-R-M2} we have considered the possibility that the
fundamental parameters $ v_{\Delta_R}$ and $v_{\delta_R}$ vary
independently. At this stage, the set of fundamental parameters is
expressed in addition to the chargino masses, reduced projectors and
eigenphases in terms of  $\tan\theta_\kappa$ and $\mu_3.$ However,
in the CP-violating case, only the phases of ${\tilde M}_L$ and
${\tilde M}_R$ have been considered as unknown, so from Eqs.
\eqref{eq:MtL-pj-etaj-thetaj} and \eqref{eq:MtR-pj-etaj-thetaj} we
get two additional equations allowing us to express
$\tan\theta_\kappa$ in terms of  the reduced projectors and
eigenphases. Since $M_R$ has been chosen real, then inserting the
above mentioned  values of $\tan\theta_\kappa$ in Eq.
\eqref{eq:MR-pj-etaj-thetaj} we get two additional equations which
can be used to determine $\mu_3$  in terms of the reduced projectors
and eigenphases. Thus, we are able to determine all the fundamental
parameters in terms of the chargino masses, the reduced projectors
and the eigenphases.

Note that in the CP-conserving case $\zeta_j=\pm 1, \forall \,
j=1,\ldots, 5,$  $\Phi_\mu =0, \pi$ and all the remaining phases are
equal to zero.  For the CP-conserving case, there are not additional
constraints to determine $\mu_3$ and $\tan\theta_k$ in terms of the
eigenphases and reduced projectors, so in order to express the
fundamental parameters in terms of them, we must know $\mu_3$ and
$\tan\theta_k$ from  other ways. In the CP-conserving case, the role
of the eigenphases is to remedy the sign ambiguity of the physical
chargino masses represented by the eigenvalues the Hermitian matrix
$H$ or ${\tilde H},$ which can be either positive or negative
\cite{key12,key13}. In addition to that, in the CP-violating case,
the eigenphases contain information about the complex phases
introduced in the chargino mass matrix.

Taking into account Eq. \eqref{eq:vDeltaRvdeltaR} and Eqs.
\eqref{eq:v-Delta-R-M1} and \eqref{eq:v-delta-R-M2}, we can express
 the phases $\Phi_1, \, \Phi_2$ and the complex parameter $\mu_3$
in terms of the chargino masses, the reduced projectors, the
eigenphases, and the fundamental parameter $\tan\theta_\upsilon.$
Indeed, let us write $p_{ij}$ and ${\tilde p}_{ij}$ in the form \be
p_{ij}= |p_{ij}| e^{i \beta_{ij}} \label{eq:pij-norma-phase} \ee and
\be {\tilde p}_{ij}= |{\tilde p}_{ij}| e^{i {\tilde
\beta}_{ij}},\label{eq:tildepij-norma-phase} \ee respectively, where
$\beta_{ij}$ and ${\tilde \beta}_{ij}$ are real phases. Eliminating
$\mu_3$ from Eqs. \eqref{eq:v-Delta-R-M1} and
\eqref{eq:v-delta-R-M2} we get \beqa \Omega_j \cos \omega_{1j} -
\Gamma_j \cos\gamma_{2j} - \Lambda_j
\cos\vartheta_j &=0,& \label{eq:Omegacos}\\
\Omega_j \sin \omega_{1j} - \Gamma_j \sin\gamma_{2j} - \Lambda_j
\sin\vartheta_j &=0,& \label{eq:Omegasin}\eeqa where

\beqa \Omega_j &=& \sqrt{2} |p_{j5}| |{\tilde p}_{j2}| M_{W_R}
\sin\theta_\upsilon \\ \Gamma_j &=& \sqrt{2} |p_{j2}| |{\tilde
p}_{j5}| M_{W_R}
\cos\theta_\upsilon, \\
\Lambda_j &=& m_{{\tilde \chi}^\pm_j} \left[ \sqrt{{P^V_{j11} \over
P^{U^\ast}_{j11}}} |p_{j5}|^2 - \sqrt{{P^{U^\ast}_{j11} \over
P^{V}_{j11}}} |{\tilde p}_{j5}|^2 \right], \eeqa

\beqa \omega_{1j} &=&  \Phi_1 +  {\tilde \beta}_{j2} - \beta_{j5}, \label{eq:omega1j}\\
\gamma_{2j} &=&  \Phi_2 + {\tilde \beta}_{j5} -
\beta_{j2}\label{eq:gamma2j}.
 \eeqa

Solving the system (\ref{eq:Omegacos}-\ref{eq:Omegasin}), we get

\be \omega_{1j} = \pm \arccos[\tau_j + \pi_j ], \qquad \gamma_{2j} =
\pm \arccos[ \nu_j + \sigma_j]\label{eq:omega1gamma2a} \ee or \be
w_{1j} = \pm \arccos[\tau_j - \pi_j ], \qquad \gamma_{2j} = \pm
\arccos[ \nu_j - \sigma_j], \label{eq:omega1gamma2b}\ee where

\be \tau_j = \left[{\Omega_j^2 - \Gamma_j^2 - \Lambda_j^2 \over 2
\Gamma_j \Lambda_j}\right] \cos\vartheta_j, \ee

\be \nu_j = \left[{\Omega_j^2 - \Gamma_j^2 + \Lambda_j^2 \over 2
\Omega_j \Lambda_j}\right] \cos\vartheta_j, \ee

\be \pi_j = \left[\sqrt{\Gamma_j^2 \, \Lambda_j^2 \,
(\Omega_j+\Gamma_j+\Lambda_j) \, (\Omega_j + \Gamma_j - \Lambda_j)
\, (\Omega_j - \Gamma_j - \Lambda_j) \, \cos^2(\vartheta_j)} \over 2
\, \Gamma_j^2 \, \Lambda_j^2 \right] \tan\vartheta_j\ee

\be \sigma_j = \left[\sqrt{\Gamma_j^2 \, \Lambda_j^2 \,
(\Omega_j+\Gamma_j+\Lambda_j) \, (\Omega_j + \Gamma_j - \Lambda_j)
\, (\Omega_j - \Gamma_j - \Lambda_j) \, \cos^2(\vartheta_j)} \over 2
\, \Omega_j \, \Gamma_j \, \Lambda_j^2 \right] \tan\vartheta_j.\ee

By comparison of Eqs.  \eqref{eq:omega1j} for $\omega_{1j}$ and
\eqref{eq:gamma2j} for $\gamma_{2j},$  with Eqs.
\eqref{eq:omega1gamma2a} and  \eqref{eq:omega1gamma2b},
respectively, we can get the phases $\Phi_1$ and $\Phi_2$ in terms
of the chargino masses, the  projector-type parameters, and
$\tan\theta_\upsilon,$ up to a eighth fold ambiguity. Finally,
inserting these results into Eqs. \eqref{eq:v-Delta-R-M1} and
\eqref{eq:v-delta-R-M2}, and regrouping the terms suitably, we get
\be \ds \mu_3=  { m_{{\tilde \chi}^\pm_j} \zeta_j \, \left[
\sqrt{{P^V_{j11} \over P^{U^\ast}_{j11}}}  + \sqrt{{P^{U^\ast}_{j11}
\over P^{V}_{j11}}} \right] - \left[{\Omega_j e^{i w_{1j}} \over
|p_{j5}|^2} + {\Gamma_j e^{i \gamma_{2j}} \over |{\tilde p}_{j5}|^2}
\right] \over \left(|p_{j5}|^2 + |{\tilde p}_{j5}|^2 \right)}.\ee In
this way, the fundamental parameter $\mu_3$ can also be expressed in
terms of the chargino masses, the reduced projectors, the
eigenphases, and $\tan\theta_\upsilon.$

In sum, we have shown that all the fundamental parameters can be
expressed in terms of the chargino masses, the projector-type
parameters,  $\tan\theta_\kappa,$ and $\tan\theta_\upsilon.$
Moreover, the considerations that follow Eq. \eqref{eq:zetaj3},
about the additional constraints in the case of the CP-violating are
also valid in this case, i.e.,  a complete disentangle of the
fundamental parameters  in terms of the chargino masses and
projector-type parameters can be reached.

\subsection{Additional set of constrains, independent reduced projector-type parameters}
The set of not trivial $V$-type reduced projectors is given by
$\{p_{j2},p_{j4}, p_{j5}\}$ and the set of
 $U^\ast$-type reduced projectors is given by  $\{{\tilde p}_{j2},{\tilde p}_{j3}, {\tilde
p}_{j5}\},$  $j=1,2,4,5.$ In the CP-violation case,  the number of
$V$-type reduced projector real parameters is $24$ and the number of
$U^\ast$-type reduced projector real parameters is also $24,$ but,
as we have seen  in Section \ref{sec-projector-formalism-uno}, the
reduced projectors are not all independent, they relate each other
by Eqs. \eqref{eq:sysV} and \eqref{eq:tsysU}.

For instance, inserting \eqref{eq:pij-norma-phase} into
\eqref{eq:sysV}, and splitting the real and imaginary part, we get
the set of $12$ constraints \be 1 + a^{(2)}_{ij} \cos(\beta_{i2}
-\beta_{j2}) + a^{(4)}_{ij} \cos(\beta_{i4} -\beta_{j4}) +
a^{(5)}_{ij} \cos(\beta_{i5} -\beta_{j5}) =0 \label{eq:syscos} \ee
and \be a^{(2)}_{ij} \sin(\beta_{i2} -\beta_{j2}) + a^{(4)}_{ij}
\sin(\beta_{i4} -\beta_{j4}) + a^{(5)}_{ij} \sin(\beta_{i5}
-\beta_{j5})=0,\label{eq:syssin}\ee where $a^{(k)}_{ij}= |p_{ik}|
|p_{jk}|, \, k=2,4,5, \, i,j=1,\ldots,5, \, ( i > j \perp i,j \ne
3).$ This means that the number of real independent parameters
describing the $V$-type reduced projectors is equal to $12.$

From Eq. \eqref{eq:sysV} we also have the following identities \be
P^V_{j11} ( 1 + |p_{j2}|^2 +  |p_{j4}|^2 + |p_{j5}|^2)= 1, \qquad
j=1,2,4,5. \ee Then, we can  parameterize the norm of the $V$-type
reduced projectors in terms of hyper-spherical angles. Indeed, we
can choose \beqa |p_{j2}| &=& \tan\psi^{(j)}
\sin\phi^{(j)} \cos\theta^{(j)},\label{eq:hyperuno} \\
|p_{j4}|&=& \tan\psi^{(j)} \sin\phi^{(j)}
\sin\theta^{(j)}, \label{eq:hyperdos}\\
|p_{j5}|&=& \tan\psi^{(j)} \cos\phi^{(j)},\label{eq:hypertres} \eeqa
with \be \sqrt{P^V_{j11}} = \cos\psi^{(j)}. \ee

In the same way, inserting \eqref{eq:tildepij-norma-phase} into
\eqref{eq:tsysU}, and splitting the real and imaginary part, we get
the set of $12$ constraints \be 1 + {\tilde a}^{(2)}_{ij}
\cos({\tilde \beta}_{i2} - {\tilde \beta}_{j2}) + {\tilde
a}^{(3)}_{ij} \cos({\tilde \beta}_{i3} - {\tilde \beta}_{j3}) +
{\tilde a}^{(5)}_{ij} \cos({\tilde \beta}_{i5} - {\tilde
\beta}_{j5}) =0 ,\label{eq:tildesyscos} \ee and \be {\tilde
a}^{(2)}_{ij} \sin({\tilde \beta}_{i2} - {\tilde \beta}_{j2}) +
{\tilde a}^{(3)}_{ij} \sin({\tilde \beta}_{i3} - {\tilde
\beta}_{j3}) + {\tilde a}^{(5)}_{ij} \sin({\tilde \beta}_{i5} -
{\tilde \beta}_{j5})=0,,\label{eq:tildesyssin}\ee where ${\tilde
a}^{(k)}_{ij}= |{\tilde p}_{ik}| |{\tilde p}_{jk}|, \, k=2,3,5, \,
i,j=1,\ldots,5, \, ( i
> j \perp i,j \ne 3).$ This means that the number of real independent parameters describing the
$U^\ast$-type reduced projectors is also equal to $12.$

Now, from Eq. \eqref{eq:tsysU} we have the following identities \be
P^{U^\ast}_{j11} ( 1 + |{\tilde p}_{j2}|^2 +  |{\tilde p}_{j3}|^2 +
|{\tilde p}_{j5}|^2)= 1, \qquad j=1,2,4,5. \ee Then, we can
parameterize the norm of the $U^\ast$-type reduced projectors in
terms of hyper-spherical angles. Indeed, we can choose \beqa
|{\tilde p}_{j2}| &=& \tan{\tilde \psi}^{(j)} \sin{\tilde \phi}^{(j)} \cos{\tilde \theta}^{(j)}, \label{eq:tildehyperuno}\\
|{\tilde p}_{j3}|&=& \tan{\tilde \psi}^{(j)} \sin{\tilde \phi}^{(j)}
\sin{\tilde \theta}^{(j)}, \label{eq:tildehyperdos}\\
|{\tilde p}_{j5}|&=& \tan{\tilde \psi}^{(j)} \cos{\tilde
\phi}^{(j)}, \label{eq:tildehypertres}\eeqa with \be
\sqrt{P^{U^\ast}_{j11}} = \cos{\tilde \psi}^{(j)}.\ee

At this stage, there are several ways of choosing the set of
independent reduced projector-type parameters. The better choice
depends on the type of problem we are analyzing and on the
experimental data. Also, it is important to consider the most
adequate set of independent parameters providing some regularities
or symmetries, allowing to solve (\ref{eq:syssin}-\ref{eq:syscos})
and (\ref{eq:tildesyssin}-\ref{eq:tildesyscos}) in an easier way.

\subsubsection{Standard set of hyper-spherical independent parameters}
Assuming  that we are able to measure the mass of the lightest
charginos and some of its cross-section related quantities, it can
be better to use the most lower index reduced projectors as
independent variables. A good choice could be the sets
$\{|p_{j2}|,|p_{j4}|,|p_{j5}|, \beta_{12},\beta_{14},\beta_{15}\},
j=1,2,4$ and $\{|{\tilde p}_{j2}|,|{\tilde p}_{j3}|,|{\tilde
p}_{j5}|, {\tilde \beta}_{12},{\tilde \beta}_{13},{\tilde
\beta}_{15}\},$ $ j=1,2,3.$

Solving the system (\ref{eq:syssin}-\ref{eq:syscos}) for the unknown
$\beta_{ij}$ variables, we get \beqa X_{ij}^{(2)} &=& {1 \over
2\,a_{ij}^{(2)}\, Z_{ij}^{(5)} } \, \Biggl\{ - \, \left( 1 +
X_{ij}^{(5)} \,a_{ij}^{(5)} \right) \,
       \left[ Z_{ij}^{(5)}  +  \left(a_{ij}^{(2)} - a_{ij}^{(4)}\right)  \left(a_{ij}^{(2)} + a_{ij}^{(4)}\right) \right]  \nonumber \\
   \nonumber \\  &\mp& {a_{ij}^{(5)}} \, \sqrt{\left( {(X_{ij}^{(5)})}^2 - 1 \right)
        \left[ Z_{ij}^{(5)}  - {\left( a_{ij}^{(2)} - a_{ij}^{(4)} \right) }^2  \right] \,
        \left[ Z_{ij}^{(5)} - {\left( a_{ij}^{(2)} + a_{ij}^{(4)} \right) }^2  \right] }\biggr\}, \label{eq:Xij2}\eeqa
\beqa  X_{ij}^{(4)}&=& {1 \over 2\, a_{ij}^{(4)} \,
    Z_{ij}^{(5)} } \, \Biggl\{ - \,  \left( 1 + X_{ij}^{(5)} \,a_{ij}^{(5)} \right) \,
       \left[ Z_{ij}^{(5)}  +  (a_{ij}^{(4)} - a_{ij}^{(2)}) (a_{ij}^{(4)} + a_{ij}^{(2)}) \right]
       \nonumber \\ & \pm  &  a_{ij}^{(5)} \, \sqrt{ \left({(X_{ij}^{(5)})}^2 - 1 \right)   \, \left[ Z_{ij}^{(5)} - {\left(
a_{ij}^{(2)} - a_{ij}^{(4)} \right) }^2  \right] \,
        \left[ Z_{ij}^{(5)} - {\left( a_{ij}^{(2)} + a_{ij}^{(4)} \right) }^2  \right] } \biggr\} ,\label{eq:Xij4}\eeqa
        where
\be X_{ij}^{(k)} = \cos(\beta_{ik} - \beta_{jk}), \ee and \be
Z_{ij}^{(5)} = \left( 1 + 2\,  X_{ij}^{(5)}  \,a_{ij}^{(5)} +
{(a_{ij}^{(5)})}^2 \right). \ee Thus from Eqs. \eqref{eq:Xij2} and
\eqref{eq:Xij4} we get \be \beta_{ik}= \beta_{1k} +
\arccos(X^{(k)}_{i1}),  \qquad i=2,4,5, \label{eq:betaik}\ee and the
constrains \beqa \arccos (X^{(k)}_{21}) + \arccos (X^{(k)}_{42}) -
\arccos (X^{(k)}_{41})&=0,& \label{eq:consuno}
\\ \arccos (X^{(k)}_{21}) + \arccos (X^{(k)}_{52}) - \arccos
(X^{(k)}_{51})&=0,& \label{eq:consdos}\\ \arccos (X^{(k)}_{42}) +
\arccos (X^{(k)}_{54}) - \arccos (X^{(k)}_{52})&=0,&
\label{eq:constres}\eeqa when $k=2,4.$

Equation \eqref{eq:betaik} allows us to express the phases
$\beta_{ik}, i=2,4,5; k=2,4$ in terms of the phases
$\beta_{12},\beta_{14}$ and $\beta_{i5}, i=1,2,4,5,$ and the norm of
the reduced projectors. On the other hand, Eq. \eqref{eq:consuno}
can be used  to determine the phases $\beta_{25}$ and $\beta_{45}$
in terms of $\beta_{15}$ and the set of independent reduced
projector norms $\{|p_{i2}|, |p_{i4}|,|p_{i5}|\}, i=1,2,4.$ Finally,
inserting the above results into Eqs. \eqref{eq:consdos} and
\eqref{eq:constres}, we get a system of equations to determine the
phase $\beta_{55}$ and the norms $\{|p_{52}|,|p_{54}|,|p_{55}|\}.$
This can be obtained in terms of the independent reduced projector
phases and norms. The same proceeding could be applied to the
treatment of the $U^\ast$-type independent reduced projector
parameters.

Thus, according to the $V$-type Eqs.
(\ref{eq:hyperuno}-\ref{eq:hypertres}), with $j=1,2,4,$ and the
analogous ones for the $U^\ast$-type, we are able to parameterize
the complete problem with six sets of hyper-spherical angles, three
$V$-type reduced projector phases and  three $U^\ast$-type reduced
projector phases.

\subsubsection{Right-handed parameters in terms of the lightest chargino parameters}
Other useful choice is to consider a set of parameters $|p_{j5}|,
\beta_{j5}$ when $j=1,2,4,$ and $|p_{5j}|, \beta_{5j},$ when $j
=2,4,5,$ expressed in terms of the reduced projector associated to
the lightest charginos, i.e., in terms of
$|p_{j2}|,|p_{j4}|,\beta_{j2},\beta_{j4},$ when $ j=1,2,4.$
Actually, it is not possible, but we can obtain a similar result
except by an independent right-handed reduced projector phase and a
dependent lightest chargino reduced projector-type parameter.

Combining Eqs.\eqref{eq:syscos} and \eqref{eq:syssin}, we get $
(i,j=1,2,4,5; i>j) $\be \beta_{i5} - \beta_{j5} = \Upsilon_{ij}
\equiv \arctan{\left(n_{ij} \over d_{ij}\right)} \label{eq:betai5}
\ee and \be a^{(5)}_{ij}= \sqrt{(n_{ij})^2 + (d_{ij})^2},
\label{eq:a5ij}\ee
 where \be n_{ij}=a^{(2)}_{ij} \sin(\beta_{i2} -\beta_{j2}) +
a^{(4)}_{ij} \sin(\beta_{i4} -\beta_{j4})\label{eq:nij} \ee and \be
d_{ij}=1 + a^{(2)}_{ij} \cos(\beta_{i2} -\beta_{j2}) + a^{(4)}_{ij}
\cos(\beta_{i4} -\beta_{j4}). \label{eq:dij} \ee Inserting Eqs.
\eqref{eq:nij} and \eqref{eq:dij} into Eq. \eqref{eq:a5ij}, we
obtain \beqa a^{(5)}_{ij} &=& \biggl\{ 1 +
\left(a^{(2)}_{ij}\right)^2 + \left(a^{(4)}_{ij}\right)^2 +  2 \,
a^{(2)}_{ij} a^{(4)}_{ij} \cos[(\beta_{i2} -\beta_{j2}) -
(\beta_{i4} - \beta_{j4})] \nonumber \\ &+&  2 \, a^{(2)}_{ij}
\cos(\beta_{i2} -\beta_{j2}) +  2 \, a^{(4)}_{ij} \cos(\beta_{i4}
-\beta_{j4})\biggr\}^{1/2}. \label{eq:a5ijexpli}\eeqa

From Eq. \eqref{eq:betai5} we get \be \beta_{i5}= \beta_{15} +
\Upsilon_{i1}, \qquad i=2,4,5, \label{eq:betai5beta15}  \ee and the
constraints \beqa \Upsilon_{21} -  \Upsilon_{41} +
\Upsilon_{42}&=&0, \label{eq:upsilonuno}\\
\Upsilon_{42} -  \Upsilon_{52} + \Upsilon_{54}&=&0,
\label{eq:upsilondos}
\\\Upsilon_{41} -  \Upsilon_{51} +
\Upsilon_{52}&=&0. \label{eq:upsilontres} \eeqa On the other hand,
combining Eqs. \eqref{eq:a5ij} for the different values of $i,j,$ we
get \beqa |p_{15}| &=& \sqrt{a^{(5)}_{51} \, a^{(5)}_{41} \over
a^{(5)}_{54} }, \qquad  |p_{25}| = \sqrt{a^{(5)}_{52} \,
a^{(5)}_{42} \over a^{(5)}_{54} }, \label{eq:normp15} \\ |p_{45}|
&=& \sqrt{a^{(5)}_{54} \, a^{(5)}_{41} \over a^{(5)}_{51} }, \qquad
|p_{55}| = \sqrt{a^{(5)}_{52} \, a^{(5)}_{51} \over a^{(5)}_{21} },
\label{eq:normp55} \eeqa  and the constrains \be a^{(5)}_{51} \,
a^{(5)}_{42}= a^{(5)}_{54} \, a^{(5)}_{21}, \qquad a^{(5)}_{51} \,
a^{(5)}_{42}= a^{(5)}_{52} \, a^{(5)}_{41}, \label{eq:a51a42}\ee
where the $a^{(5)}_{ij}$ factors are given by the right side member
of Eq. \eqref{eq:a5ijexpli}.

In Eq. \eqref{eq:betai5beta15} the phases $\beta_{i5}, i=2,4,5,$ are
expressed in terms of $\beta_{15}$ and the parameters $|p_{i2}|,$
$|p_{i4}|,$ $\beta_{i2},$ $\beta_{i4},$ $ i=1,2,4,5.$ Similarly,
Eqs. \eqref{eq:normp15} and \eqref{eq:normp55} imply that $|p_{i5}|,
\, i=1,2,4,5,$ can be expressed in terms of the parameters
$|p_{i2}|, |p_{i4}|,\beta_{i2},\beta_{i4}, \, i=1,2,4,5.$  However,
Eqs. \eqref{eq:upsilondos} and \eqref{eq:upsilontres} together with
the constraints \eqref{eq:a51a42}, constitute a system of four
algebraic equations relating the parameters $|p_{52}|,|p_{54}|,
\beta_{52}, \beta_{54},$ to $|p_{j2}|,|p_{j4}|, \beta_{j2},
\beta_{j4},$ when $j=1,2,4.$ Thus, in principle, we can use these
equations to express the set of four parameters in terms of the last
set of parameters. Moreover, the constraint given in Eq.
\eqref{eq:upsilonuno}, only involves the $|p_{j2}|,|p_{j4}|,
\beta_{j2}, \beta_{j4},$ when $j=1,2,4.$ Thus, that equation can be
used to express one of the parameters of the set in terms of the
rest of the parameters of the same set. In sum, the set of $12$
independent parameters can be chosen to be $\beta_{15}$ plus  $11$
parameters taken  from $|p_{j2}|,|p_{j4}|, \beta_{j2}, \beta_{j4},$
when $j=1,2,4.$

A similar way can be used for $U^\ast$-type parameters.

\section{Conclusion}
\label{sec-conclusion} In this paper we have computed analytically
the chargino mass spectrum, at the tree level, in the context of the
LRSUSY model, including a general set of CP-violating phases,
$\Phi_L, {\tilde \Phi}_L, {\tilde \Phi}_R, \Phi_\mu, \Phi_1, \Phi_2,
\Phi_3$.

We have shown that there is always a neighborhood in the fundamental
parameter space where one of the chargino masses relates in a simple
manner with the parameters $|\mu|.$ This fact allows us  to
factorize the quintic polynomial representing the characteristic
equation, which is used to determine the chargino mass spectrum, and
to arrange the charginos in a determined neighborhood according to
the size of their masses. We have also shown that in the most
general CP-violating scenario the chargino masses depend only on
three global phases $\Theta_1, \Theta_2,$ and $\Theta_3.$

We have computed analytically the diagonalizing matrices $V$ and
$U^\ast$. The entries of these matrices can be expressed in terms of
the more fundamental quantities $\Delta^{(\ell)}_{ij}$ and ${\tilde
\Delta}^{(\ell)}_{ij},$ respectively. Then, with the help of these
fundamental quantities we have implemented a generalized projector
formalism which provides a system of basic equations connecting the
reduced projectors, the eigenphases and the chargino masses with the
chargino parameters. These equations constitute the keystone on
which the parameter inversion process is based. Some connections
with the Jarlskog's formulation allows us to disentangle in a direct
manner some relevant parameters, specially those ones lying on the
diagonal of the chargino mass matrix.

We have  shown that a systematic reconstruction of the fundamental
parameter is possible if we are able to measure an adequate set of
observables, for instance, some cross-section type observables
derived from the chargino pair production in electron-positron
annihilation.

Concerning this last point, we have shown that it is possible to
re-parameterize the cross-section type observables in terms of the
chargino masses, the reduced projectors, and the eigenphases. The
minimal number of reduced projector-type parameters that we can use
to parameterize this kind of observables is $24,$ that is, $12$ for
$V$-type and $12$ for $U^\ast$-type.  We have seen that there are
many ways to choose the set of independent reduced projector-type
parameters, we have analyzed two of them. The first one consists  of
six sets of hyper-spherical angles, three $V$-type reduced projector
phases and three $U^\ast$-type reduced projector phases. The second
one involves eleven $V$-type  and eleven $U^\ast$-type reduced
projector parameters associated to the lightest charginos, and one
$V$-type and one $U^\ast$-type reduced projector phases associated
to the right-handed contribution.

The systematic inversion method used in this paper to determine the
fundamental chargino parameters, based on measurements of physical
observables, can be applied to any number of charginos and to any
number of neutralinos, no matter the particular model we used to
describe them.
\section*{Acknowledgments} N. Alvarez Moraga thank all his family for
important encouragement and valuable support.
\appendix
\section{\boldmath $H$ and $\tilde H$'s  entries}
\label{sec-helements} The matrices $H$ and $\tilde H$ in terms of
the entries of the mixing matrix $M$ are given by $H_{ij}=
\sum_{k=1}^5 M^{\ast}_{ki} M_{kj}$ and ${\tilde H}_{ij}=
\sum_{k=1}^5 M_{ik} M^{\ast}_{jk}, $ respectively. We get that
\beqa \nonumber H_{11}&=&  |M_L|^2 + 2 |{\tilde M}_L|^2 \sin^2\theta_\kappa, \nonumber \\
\nonumber H_{22} &=&  M_{R}^2 + 2  |{\tilde M}_R|^2
\sin^2\theta_\kappa + g^2_R \, (v_{\delta_R})^2, \nonumber \\
\nonumber H_{33}&=& |\mu|^2, \nonumber \\ \nonumber H_{44}&=&
|\mu|^2 + 2
(|{\tilde M}_L|^2 + |{\tilde M}_R|^2) \cos^2\theta_\kappa, \nonumber \\
\nonumber H_{55}&=& g^2_R \, (v_{\Delta_R})^2 + |\mu_3|^2, \nonumber
\\\nonumber H_{12}&=& H_{21}^\ast =  2 |{\tilde M}_L|  |{\tilde
M}_R| e^{i ({\tilde \Phi}_R - {\tilde \Phi}_L)} \sin^2\theta_\kappa,
\nonumber \\
\nonumber H_{13}&=& H_{31}^\ast =  0,
\nonumber \\
\nonumber H_{14}&=& H_{41}^\ast = \sqrt{2}|{\tilde M}_L| \bigl[e^{i
({\tilde \Phi}_L - \Phi_L)} |M_L| \cos\theta_\kappa  -  e^{i (
\Phi_\mu - {\tilde \Phi}_L)} |\mu| \sin\theta_\kappa\bigr],
\nonumber \\
\nonumber H_{15}&=& H_{51}^\ast =  0,
\nonumber \\
\nonumber H_{23}&=& H_{32}^\ast = 0,
\nonumber \\
\nonumber H_{24}&=& H_{42}^\ast = \sqrt{2}|{\tilde M}_R| \bigl[e^{i
{\tilde \Phi}_R } M_R  \cos\theta_\kappa -  e^{i ( \Phi_\mu -
{\tilde \Phi}_R)} |\mu| \sin\theta_\kappa\bigr],
\nonumber \\
\nonumber H_{25}&=& H_{52}^\ast = g_R [ M_R \, v_{\Delta_R} e^{i
\Phi_1} + v_{\delta_R} \, |\mu_3| e^{i (\Phi_3 - \Phi_2)} ], \nonumber \\
\nonumber H_{34}&=& H_{43}^\ast = 0, \nonumber \\ \nonumber
H_{35}&=& H_{53}^\ast = 0,
\nonumber \\ \nonumber H_{45}&=& H_{54}^\ast = \sqrt{2} g_R  \, {\tilde M}_R \, v_{\Delta_R} \, e^{i (\Phi_1 - {\tilde \Phi}_R)} \cos\theta_\kappa , \nonumber \\
\label{eq:Hij} \eeqa and  ${\tilde H}_{ij}= \sum_{k=1}^5 M_{ik}
M^{\ast}_{jk}
:$ \beqa \nonumber {\tilde H}_{11}&=&  |M_L|^2 + 2 |{\tilde M}_L|^2 \cos^2\theta_\kappa, \nonumber \\
\nonumber {\tilde H}_{22} &=&  M_{R}^2 + 2  |{\tilde M}_R|^2
\cos^2\theta_\kappa + g^2_R \, (v_{\Delta_R})^2 , \nonumber \\
\nonumber {\tilde H}_{33}&=& |\mu|^2 + 2 (|{\tilde M}_L|^2 +
|{\tilde M}_R|^2) \sin^2\theta_\kappa, \nonumber \\ \nonumber
{\tilde H}_{44}&=& |\mu|^2, \nonumber \\ \nonumber {\tilde
H}_{55}&=&  g^2_R \, (v_{\delta_R})^2 + |\mu_3|^2, \nonumber
\\\nonumber {\tilde H}_{12}&=& {\tilde H}_{21}^\ast =  2 |{\tilde
M}_L| |{\tilde M}_R| e^{i ({\tilde \Phi}_L - {\tilde \Phi}_R)}
\cos^2\theta_\kappa,
\nonumber \\
\nonumber {\tilde H}_{13}&=& {\tilde H}_{31}^\ast = \sqrt{2}|{\tilde
M}_L| \bigl[e^{i (\Phi_L - {\tilde \Phi}_L )} |M_L|
\sin\theta_\kappa  - e^{i ({\tilde \Phi}_L - \Phi_\mu)} |\mu|
\cos\theta_\kappa\bigr],\nonumber \\
\nonumber {\tilde H}_{14}&=& {\tilde H}_{41}^\ast =  0, \nonumber \\
\nonumber {\tilde H}_{15}&=& {\tilde H}_{51}^\ast =  0,
\nonumber \\
\nonumber {\tilde H}_{23}&=& {\tilde H}_{32}^\ast = \sqrt{2}|{\tilde
M}_R| \bigl[e^{- i {\tilde \Phi}_R } M_R  \sin\theta_\kappa  - e^{i
({\tilde \Phi}_R - \Phi_\mu  )} |\mu| \cos\theta_\kappa\bigr],
\nonumber \\
\nonumber {\tilde H}_{24}&=& {\tilde H}_{42}^\ast = 0, \nonumber \nonumber \\
{\tilde H}_{25}&=& {\tilde H}_{52}^\ast = g_R \, [v_{\Delta_R} \,
|\mu_3|
e^{i( \Phi_1 - \Phi_3)} + M_R \, v_{\delta_R} e^{-i \Phi_2}], \nonumber \\
\nonumber {\tilde H}_{34}&=& {\tilde H}_{43}^\ast = 0, \nonumber \\
\nonumber {\tilde H}_{35}&=& {\tilde H}_{53}^\ast = \sqrt{2} \, g_R
\, |{\tilde M}_R| v_{\delta_R}\, e^{i ({\tilde \Phi}_R - \Phi_2)} \,
\sin\theta_\kappa,
 \nonumber \\
\nonumber {\tilde H}_{45}&=& {\tilde H}_{54}^\ast =  0,\nonumber \\
\label{eq:THij} \eeqa where $v_{\Delta_R}$ and $v_{\delta_R}$ are
defined  in Eq. \eqref{eq:vDeltaRvdeltaR}.

\end{document}